\title{Full Abstraction for a Recursively Typed\\
Lambda Calculus with Parallel Conditional
\thanks{revised version of Report 12/1993 of SFB 124, FB 14 - Informatik,
Universit\"at des Saarlandes, Saarbr\"ucken 1993}}
\author{Fritz M\"uller\\
email: \texttt{($\lambda$x.muellerxcs.uni-sb.de)@}\\
URL: \texttt{rw4.cs.uni-sb.de/\~{}mueller/recursive.ps.gz}}
\date{1993}
\documentclass[12pt,fleqn,leqno]{article}
\usepackage{latexsym}
\usepackage{epic,eepic}
\newcommand{\dl}{\drawline}
\mathindent=\parindent
	\newcommand{\ie}{i.e.\ }
	\newcommand{\eg}{e.g.\ }
	\newcommand{\wrt}{w.r.t.\ }

	\newcommand {\ih}{induction hypothesis\ }
	\let \3=\ss
	\newcommand{\LP}{\par\noindent}
	\newcommand{\sma}{\smallskipamount}
 	\newcommand{\med}{\medskipamount}
        \newcommand{\de}{\em}         %definierte Woerter
        \newcommand{\pr}{\sf}         %Programm-Woerter
        \newcommand{\pro}[1]{\mbox{\pr #1}}  %Programme
        \newcommand{\CA}[1]{\mbox{$\cal{#1}$}}
        \newcommand{\Gm }{\sym{\mu          }}

        \newcommand{\Gt }{\sym{\tau         }}
        
	\newcommand{\GO }{\sym{\Omega       }}
        \newcommand{\sym}[1]{\ifmmode \mathord{#1} \else$#1$ \fi}
	\newcommand{\symrel}[1]{\ifmmode \mathrel{#1} \else$#1$ \fi}
        
	\newlength{\arrow}
        \newcommand{\Ra}{\;\Rightarrow\;}
        \newcommand{\La}{\;\Leftarrow\;}
	\newcommand{\Lra}{\;\Leftrightarrow\;}
        \renewcommand{\=}{=_{\mbox{\scriptsize def}}} 
        \newcommand{\sub}{\subseteq}
 	\renewcommand{\sup}{\supseteq}
	\newcommand{\suq}{\sqsubseteq}

	\newcommand{\mb}{\makebox(0,0)}
\newtheorem{theorem}{Theorem}[section]
\newtheorem{definition}[theorem]{Definition}
\newtheorem{lemma}[theorem]{Lemma}
\newtheorem{proposition}[theorem]{Proposition}
\newtheorem{prop}[theorem]{Proposition}
\newtheorem{corollary}[theorem]{Corollary}

\newcommand {\THEOREM}{\begin{theorem}}
\newcommand {\ETHEOREM}{\end{theorem}}
\newenvironment{defi}{\begin{definition}\rm}{\end{definition}}
\newcommand {\DEFI}{\begin{defi}}
\newcommand {\EDEFI}{\end{defi}}
\newcommand {\LEMMA}{\begin{lemma}}
\newcommand {\ELEMMA}{\end{lemma}}
\newcommand {\PROP}{\begin{prop}}
\newcommand {\EPROP}{\end{prop}}

\newcommand {\EQN}{\begin{eqnarray*}}
\newcommand {\EEQN}{\end{eqnarray*}}
\newcommand {\ARR}{\[\begin{array}}
\newcommand {\EARR}{\end{array}\]}

\newcommand{\proof}{\noindent{\bf Proof:}\ }
\newcommand{\eproof}{\hfill \rule{3mm}{3mm}}

\newenvironment{deflist}[1]%
{\begin{list}{}%
{\settowidth{\labelwidth}{#1}%
\setlength{\leftmargin}{\labelwidth}%
\addtolength{\leftmargin}{\labelsep}%
}}%
{\end{list}}

\newenvironment{deflisteng}[1]%
{\begin{list}{}%
{\settowidth{\labelwidth}{#1}%
\setlength{\leftmargin}{\labelwidth}%
\addtolength{\leftmargin}{\labelsep}%
\setlength{\topsep}{0pt}%
\setlength{\itemsep}{0pt}%
}}%
{\end{list}}%

\newcommand {\DL}{\begin{deflist}}
\newcommand {\EDL}{\end{deflist}}
\newcommand {\DLE}{\begin{deflisteng}}
\newcommand {\EDLE}{\end{deflisteng}}
\renewcommand{\l}{\lambda}
\renewcommand{\t}{\tau}
\newcommand {\s}{\sigma}
\renewcommand {\r}{\varrho}
\newcommand {\m}{\mu}
\newcommand {\ph}{\varphi}

\renewcommand {\O}{\Omega}

\newcommand {\f}{\rightarrow}
\newcommand {\lf}{\leftarrow}
\newcommand {\fR}{\rightarrow_{\cal R}}
\newcommand {\rast}{\rightarrow^{\ast}}
\newcommand {\last}{\leftarrow^{\ast}}
\newcommand {\fu}{\symrel{\rightarrow}}
\newcommand {\fus}{\mbox{$\rightarrow$}}

\newcommand {\subst}{\sym{:=}}

\newcommand {\sconst}[1]{\mbox{\it #1}}
\newcommand {\lift}{\pro{lift}}
\newcommand {\void}{\pro{void}}
\newcommand {\bool}{\pro{bool}}

\newcommand {\cond}{\pro{case}}
\newcommand {\case}{\pro{case}}
\newcommand {\scase}{\sconst{case}}
\newcommand {\pcond}{\pro{pcase}}
\newcommand {\pcase}{\pro{pcase}}
\newcommand {\spcase}{\sconst{pcase}}
\newcommand {\pair}{\pro{pair}}
\newcommand {\spair}{\sconst{pair}}
\newcommand {\fst}{\pro{fst}}
\newcommand {\sfst}{\sconst{fst}}
\newcommand {\snd}{\pro{snd}}
\newcommand {\ssnd}{\sconst{snd}}
\newcommand {\0}{\mbox{\pr 0}}
\newcommand {\1}{\mbox{\pr 1}}
\newcommand {\Outz}{\sconst{Out0}}
\newcommand {\Outi}{\sconst{Out1}}
\newcommand {\outz}{\pro{out0}}
\newcommand {\outi}{\pro{out1}}
\newcommand {\andf}{\pro{and}}
\newcommand {\ifz}{\pro{if}}
\newcommand {\notf}{\pro{not}}
\newcommand {\orf}{\pro{or}}
\newcommand {\pcf}{\pro{pc}}
\newcommand {\sbf}{\pro{sb}}

\newcommand {\Tm}{T_{\m}}
\newcommand {\Tmc}{T_{\m}^{c}}

\newcommand {\Tinf}{T_{\infty}}
\newcommand {\Term}{{\cal T}}
\newcommand {\T}{{\cal T}}     
\newcommand {\Tc}[1]{{\cal T}^c_{\scriptsize #1}}
\newcommand {\Prog}{\sconst{Prog}}
\newcommand {\N}{{\cal N}}        
\newcommand {\as}{^{\ast}}
\newcommand {\LR}{\langle L \mathord{\rightarrow} R \rangle}
\newcommand {\ST}{\langle S \mathord{\rightarrow} T \rangle}
\newcommand {\p}{\times}
\newcommand {\cp}{\times} %matematical
\newcommand {\sust}{\unlhd}
\newcommand {\pre}{\prec}
\newcommand {\Sem}{{\cal S}}
\renewcommand {\S}[1]{\Sem[\![#1]\!]}
\newcommand {\SH}[1]{\Sem^H[\![#1]\!]}
\newcommand {\Sb}[1]{\Sem[\![#1]\!]\bot}
\newcommand {\ep}{\varepsilon}  % environment
\newcommand {\Se}[1]{\Sem[\![#1]\!]\ep}
\newcommand {\Sae}[1]{\overline{\Sem}[\![#1]\!]\ep}
\newcommand {\Sa}[1]{\overline{\Sem}[\![#1]\!]}
\newcommand {\Env}{\sconst{Env}}
\newcommand {\mto}{\mapsto}
\newcommand {\Oper}{{\cal O}}
\newcommand {\Op}[1]{{\cal O}[\![#1]\!]}
\newcommand {\A}[1]{{\cal A}(#1)}
\newcommand {\B}[1]{{\cal B}(#1)}
\newcommand {\C}[1]{{\cal C}(#1)}
\newcommand {\pj}{\sqcup}
\newcommand {\tc}{\mathrel{@}}
\newcommand {\unfold}{\leadsto}
\newcommand {\unfolds}{\leadsto^{\ast}}
\newcommand {\PSys}{\mbox{\bf PSys}}
\newcommand {\ps}[1]{\mbox{$\cal #1$}}
\newcommand {\psA}{\ps{A}}
\newcommand {\psB}{\ps{B}}
\newcommand {\con}{\uparrow}
\newcommand {\lep}{\leq}
\newcommand {\gep}{\geq}
\newcommand {\downc}[1]{#1\!\! \downarrow}
\newcommand {\ele}[1]{|#1|}
\newcommand {\eleps}[1]{\ele{\ps{#1}}}
\newcommand {\elepsA}{\eleps{A}}
\newcommand {\elepsB}{\eleps{B}}
\newcommand {\ubot}{\underline{\bot}}
\newcommand {\Achain}{\psA_0 \sust \psA_1 \sust \ldots}
\newcommand {\Ai}{\psA_i = (A_i,\con_i,\lep_i)}
\newcommand {\cupAi}{\bigcup_i \psA_i}
\newcommand {\st}{\: | \:}
\renewcommand {\.}{.\;}
\newcommand {\e}{\:\wedge\:}
\newcommand {\et}{\mbox{\ and\ }}
\newcommand {\set}[1]{\{ #1 \}}
\newcommand {\ov}[1]{\overline{#1}}
\newcommand {\ovA}{\ov{A}}
\newcommand {\fl}[1]{\ele{#1}}
\newcommand {\flr}{\fl{r}}
\renewcommand {\Pr}{\sconst{Pr}}
\renewcommand {\P}{{\cal P}}
\newcommand {\level}{\sconst{level}}
\newcommand {\dap}{\lhd}
\newcommand {\ap}{\mathrel{\mbox{$\lhd\!\!\!\!\!-$}}}
\newcommand {\apind}{\mbox{\scriptsize $\lhd\!\!\!\!\!-$}}
\newcommand {\ip}[1]{<_\ep^{#1}}
\newcommand {\ips}{\ip{\s}}
\newcommand {\ipt}{\ip{\t}}
\newcommand {\ipr}{\ip{\r}}
\newcommand {\ipp}{\ip{\psi}}
\newcommand {\ipe}[2]{<_{#1}^{#2}}
\newcommand {\BU}{$\bullet$\ }
\newcommand {\tild}[1]{\overline{#1}}
\newcommand {\tile}{\tild{\ep}}

\newcommand {\Cond}{\sconst{Cond}}
\newcommand {\term}{\mathrel{\sconst{term}}}
\newcommand {\eq}{\mathrel{\sconst{eq}}}
\newcommand {\ncon}{\mathrel{\mbox{$\con\!\!\!\!\!-$}}}
\newcommand {\co}{\mathord{\f}}

\newcommand {\te}{\theta}
\newcommand {\ovte}{\overline{\te}}
\newcommand {\Pcase}{\sconst{Pcase}}
\newcommand {\sbool}{\scriptsize \bool}
\newcommand {\app}{\sconst{app}}
\begin{document}
\maketitle
\begin{abstract}
We define the syntax and reduction relation of a recursively typed lambda
calculus with a parallel case-function (a parallel conditional).
The reduction is shown to be confluent.
We interpret the recursive types as information systems in a restricted form,
which we call {\em prime systems}. A denotational semantics is defined
with this interpretation. We define the syntactical normal form approximations
of a term and prove the Approximation Theorem: The semantics of a term
equals the limit of the semantics of its approximations. The proof uses
inclusive predicates (logical relations).
The semantics is adequate with respect to the
observation of Boolean values. It is also fully abstract in the presence
of the parallel case-function.

\end{abstract}

\section{Introduction}

In his seminal paper \cite{Plotkin}, Gordon Plotkin explores the
relationship between the operational (reduction) semantics and the denotational
semantics of the functional programming language PCF. PCF is a call-by-name
typed lambda calculus with the ground types boolean and integer,
and any functional type.
In order to compare operational and denotational semantics, one defines a notion
of operational observation and a preorder on terms induced by this notion.
In the case of PCF, the observation is of integer values only, and the preorder
is defined by observation of arbitrary terms through integer contexts.
The closed terms of ground type integer are singled out as {\em programs}.
Programs are regarded as the only terms whose syntactical values (integers)
can be observed directly. If the  
semantics of a program $M$ is an integer
value $i$, then $M$ can be reduced to $i$. This result is called the
{\em adequacy} of the semantics.
(The denotational semantics is simply called the semantics here and
in the following.)

A more general result about terms of any type is the {\em Approximation
Theorem} or limiting completeness, as proved in \cite{Wadsworth} for
the untyped lambda calculus and in \cite{Berry} for PCF.
The approximations of a term $M$ are defined, roughly, as the normal
form prefixes of the reducts of $M$. The Approximation Theorem states
that the semantics of a term equals the limit of the semantics of its
approximations.

Plotkin's programme proceeds as follows: The operational preorder on terms
is defined as $M \suq N$ iff for all contexts $C[\:]$ such that $C[M]$ and
$C[N]$ are programs: if $C[M]$ reduces to a value $i$, then also $C[N]$.
If $\S{M} \suq \S{N}$, where $\Sem$ is the semantics function, then
$M \suq N$; this follows from adequacy. The converse, if $M \suq N$
then $\S{M} \suq \S{N}$, is not true for PCF with only sequential
operations. This is due to the fact that there are parallel functions in the
semantic model, like the parallel or, that cannot be defined syntactically.
But when a parallel if-operation, or the parallel or, is added to the
syntax, then ``if $M \suq N$ then $\S{M} \suq \S{N}$'' holds.
This is called the {\em full abstraction} of the semantics;
the operational and denotational preorders on terms coincide.

We elaborate the programme above for a call-by-name recursively typed
lambda calculus and establish similar results : 
Approximation Theorem and adequacy for the sequential or parallel calculus
and full abstraction for the parallel calculus only.

Chapter 2 defines the syntax and the reduction relation of our calculus.
Types are built up from the separated sum +, the cartesian separated product
$\p$, the function space $\f$, and recursion. Every recursive type denotes a
possibly infinite type tree. Recursive types with the same type tree
are regarded as equivalent. 
Terms are built up from variables, $\l$-abstraction, application, and
constants for the type constructors + and $\p$. Among the constants is
a parallel case operation \pcase. The operational semantics is defined by
the one-step reduction $\f$ of a redex in any context. We prove that
reduction is confluent. For the proof we use the confluence theorem of
\cite{Mueller:lconfluence} which says roughly: The combination of
the lambda calculus with a confluent, left-linear and not variable-applying
algebraic term rewriting system is confluent.

The subsequent chapters explore the semantics. We use information systems
to give the semantics of recursive types \cite{Larsen/Winskel,Winskel:book}.
Chapter 3 introduces a specialized form of information systems that we call
{\em prime systems}: Here the predicates of consistency and entailment
are given by binary relations on the set of {\em primes} (= tokens).
Prime systems were first introduced for different purposes under the name
event structures in \cite{NielsenPW} and shown to be equivalent to
prime algebraic coherent partial orders.
We transfer the results of \cite{Larsen/Winskel} to our prime systems:
The class of prime systems is a complete partial order under the
substructure relation $\sust$. We define operations on prime systems
corresponding to our type constructors +, $\p$, $\f$ and show that they
are continuous.

This enables us, in Chapter 4, to give a semantic interpretation
of type trees and recursive types as prime systems.
The interpretation of finite prefixes of a type tree gives a $\sust$-chain
of prime systems; the interpretation of the whole type tree is the limit
of this chain. Note that the primes at one level of the chain are
directly contained in the following levels; there is no need for 
embedding-projection pairs as in the inverse limit solution of recursive
domain equations. This is an advantage of the concrete representation
of domains by information systems or prime systems. Anyway, this
concrete representation of domain elements by sets of primes will be needed
to prove full abstraction. 
Chapter 4 also gives the semantics function $\Sem$ on terms and proves
its soundness: Reduction does not change the semantics of terms.

Chapter 5 proves the Approximation Theorem. We define a prefix order
$\pre$ on terms where the constant $\O$ is the least term.
A normal form $A$ is an {\em approximation} of a term $M$ iff there is
a reduct $N$ of $M$ such that $A \pre N'$ for all reducts $N'$ of $N$.
The set $\A{M}$ of approximations of $M$ is an ideal and can be seen
as the syntactic value or B\"ohm tree of $M$.
For the parallel calculus,
it is not possible to define approximations
by an analogue of head normal forms. But for the sequential calculus
(without \pcase), we give two analogues of head normal forms to define
alternative sets of approximations.
The Approximation Theorem says that the semantics of a term equals
the limit of the semantics of its approximations.
This is proved by the inclusive predicate technique, as it was used
in \cite{Mosses/Plotkin} to prove the analogous theorem for the
untyped lambda calculus. We adapt the technique to prime systems:
We give an inductive definition of the inclusive predicates (logical
relations) on the
primes of our prime system interpretation of types.

Chapter 6 proves adequacy and full abstraction of the semantics.
We have to define a notion of observation and the corresponding operational
preorder on terms. We choose to observe the values \0\ and \1\ of type
$\bool = \void + \void$, where \void\ is the type of just one bottom
element. So our {\em programs} are the closed terms of type \bool.
For a program $M$ we define the operational value $\Op{M}$ as 0 or 1      
if $M$ reduces to \0\ or  \1\, respectively, and as $\bot$ otherwise.
The Adequacy Theorem says that $\Op{M} = \Sb{M}$ for every program $M$;
it is a consequence of the Approximation Theorem.

The operational preorder on terms is defined as $M \suq N$ iff for all
contexts $C[\:]$ such that $C[M]$ and $C[N]$ are programs,
$\Op{C[M]} \sub \Op{C[N]}$ holds.
Again we have: If $\S{M} \sub \S{N}$, then $M \suq N$, as a consequence
of adequacy.
Full abstraction, $M \suq N$ iff $\S{M} \sub \S{N}$, is proved for
the parallel calculus.
As in \cite{Plotkin} the proof is based on the Definability Lemma:
For all finite elements $d$ of a semantic domain there is a term $M$
with $\Sb{M} = d$. The proof uses the representation of elements as
sets of primes.

The last Chapter 7 proves that the \pcase-function is definable from
the parallel \andf\ function.

\subsection*{Related work}

Recently, \cite{Winskel:book} gave two recursively typed $\l$-calculi
with their denotational semantics, by information systems, and proved
the adequacy by the inclusive predicate (logical relation) technique.
The first calculus has an eager (call-by-value) operational semantics.
The second one has lazy (call-by-name) operational semantics like ours,
but a different notion of observation is chosen:
For every type certain terms are singled out as {\em canonical forms}.
For product types these are the terms $(M,N)$, for sum types 
$\pro{inl}(M)$ and $\pro{inr}(M)$, and for function types the terms
$\l x.M$.
The observation that is made of terms is the convergence to a canonical
form. The given denotational semantics is adequate with respect to this
notion of observation.
This means that a term converges to a canonical form iff its semantics
is not bottom.
Especially, the semantics of every term $\l x.M$ is not bottom,
whereas we have $\Se{\l x.\O}=\bot$.

Finally some remarks on coalesced sums and the observation of
termination for all types. We did not include the coalesced sum in our
type system, only separated sums. The coalesced sum of two domains
is the disjoint union of the domains, with the two bottom elements identified.
A coalesced sum would demand {\em strict} constructors
{\pr inl}: $\t \f \t\oplus \r$ and {\pr inr}: $\r \f \t \oplus \r$.
 These constructors have to evaluate their arguments to a non-bottom value
before they can be used by a \case-operation. (In contrast our corresponding
constructors \0\ and \1\ are non-strict; they can be used without evaluated
argument.) But the detection of non-bottom values is a complicated task
for functional types, when we assume our denotational semantics of functions.
On the other side I see no use for coalesced sums of functional types.
Therefore I think that coalesced sums should be restricted to
non-functional types, so that \eg the recursive definition of the flat cpo
of integers becomes possible. The check for non-bottomness of functional
values, if it is desired, should be programmed using special functions
incorporated in the language, \eg Plotkin's ``exists'' operator.

\cite{Cosmadakis} constructs evaluators for a
recursively typed lambda calculus with coalesced sums and strict, coalesced
products of any type. The notion of observation for these evaluators is
the observation of termination for terms of all types. The relation of
operational and denotational semantics is given by the property
of ``complete adequacy'': The semantics of any term is non-bottom iff
its evaluation terminates. This ensures the detection of non-bottomness
for coalesced sums. 
The work succeeds with a trick: The semantic domains are lattices;
top elements (that are not syntactically definable) are added to the
domains. Thus a term like $\l x.\pro{if}\: x \: (\pro{if}\: x \: \O\: \0)\: \O$,
whose normal
semantics is $\bot$, now becomes non-bottom. For the normal cpo semantics
only a vague sketch of an evaluator is given.

There has been later work proving adequacy for a lazy functional language
with recursive and \emph{polymorphic} types, also using information systems
\cite{Blaaberg}.

\section{Syntax and reduction}

\subsection{Types}

We adopt the syntax of the recursive type system of 
\cite{Cardone/Coppo,Cardone/CoppoRec}.
Especially, recursive types are considered equivalent
if they have the same unfoldings as regular trees. But instead of type
constants we have some more type constructors besides \fus.
The {\de type expressions}
are given by the following grammar, where $t$ stands for
elements of a denumerable set $V_T$ of type variables:
\[ \Gt ::= t \mid  \Gt + \Gt \mid \Gt \times \Gt \mid
\Gt \fu \Gt \mid \Gm t.\Gt \mid \void \]
$\Tm$ is the set of all type expressions.
$\Tmc$ is the set of all closed type expressions, called {\de types}.

\LP
We give the informal meaning of types in terms of domains:
\begin {deflisteng}{$\s \f \t$}
\item[$\s + \t$]   is the separated sum of $\s$ and $\t$,
\item[$\s \times \t$] is the cartesian separated product of $\s$ and $\t$,
\item[$\s \f \t$] is the space of continuous functions from $\s$ to $\t$,
\item[$\m t.\t$]    is the fixed point of the mapping $t \mapsto \t$,
 	           the solution of the recursive domain equation $t = \t$,
\item[\void]       is the canonical notation of the undefined type;
		   it has the same meaning as $\m t.t$. 
		   In \cite{Cardone/Coppo} it is called $\O$. 
                   The corresponding domain
		   has just one element $\bot$.
\end{deflisteng}

\LP
We define the {\de simple types} by the grammar:
\[ \t ::= \void \mid \t + \t \mid \t \p \t \mid \t \f \t \]
$T$ is the set of all simple types. It is $T \sub \Tmc$.

\DEFI
The \void-prefix order $\pre\; \sub T \cp \Tmc$ is the least partial order
satisfying:
\begin{deflisteng}{1)}
\item[1)] $\void \pre \t$ for all $\t \in \Tmc$,
\item[2)] $\s \pre\s', \t \pre \t' \Ra \s \tc \t \pre \s' \tc \t'$\\
          for $\tc\: \in \set{ +,\p,\f}$, $\s,\t\in T$, and $\s',\t'\in \Tmc$.
\end{deflisteng}
\EDEFI

$\pre$ is a partial order on $T$. For every $\s,\t \in T$ with an upper
bound there is a least
upper bound $\s \pj \t \in T$. 
$\Tinf$ denotes the ideal completion of $T$, \ie the set of ideals of 
simple types,
ordered by $\sub$. Here ideals are sets $I$ of simple types that are
non-empty,
downward closed: $\t \in I \e \s \pre \t \Ra \s \in I$, and directed:
for all $\s,\t\in I$ there is $\r\in I$ with $\s\pre\r$ and $\t\pre\r$.
The elements of $\Tinf$ are called {\de type trees} and are also denoted
by $\s,\t,\r$.

\LP
We define $\void\in \Tinf$ as $\void = \set{\void}$. For $\tc\: = +,\p,\f$
and $\s,\t \in \Tinf$ we define
\[ \s \: \tc\: \t = \{\void\}
\cup \{\s'\: \tc\: \t' \mid \s'\in \s \e \t'\in \t\} \]
Every type tree of $\Tinf$ has one of the forms
\void, $\s +\t$, $\s\p\t$, $\s\f\t$ with unique $\s,\t \in \Tinf$.

\DEFI
The {\de unfolding} $\unfold\; \sub \Tmc \cp \Tmc$ is the least relation
satisfying:
\begin{deflisteng}{1)}
\item[1)]
$\m t.\t \unfold \t[\m t.\t / t]$\\
The right term is the replacement of $\m t.\t$ for all free occurrences
of $t$ in $\t$; it is also closed. Note that $\m t.\t$ does not contain free
variables that could be bound after the replacement.
\item[2)]
$\t \unfold \t' \Ra (\t \tc \s) \unfold (\t' \tc \s)$
and $(\s \tc \t) \unfold (\s \tc \t')$
for $\tc\: \in \set{ +,\p,\f}$, $\t,\t',\s \in \Tmc$.
\end{deflisteng}
\EDEFI

$\unfold$ reduces only one {\em outermost} redex $\m t.\t$.
The outermost redexes are disjoint, therefore $\unfold$ fulfills the
diamond property: If $\t \unfold \s$ and $\t \unfold \r$, then there is $\psi$
with $\s \unfold \psi$ and $\r \unfold \psi$.

$\unfolds$ is the reflexive, transitive closure of $\unfold$.
It is confluent: If $\t \unfolds \s$ and $\t \unfolds \r$,
then there is $\psi$ with $\s \unfolds \psi$ and $\r \unfolds \psi$.\\
If $\s \pre \t$ and $\t \unfolds \t'$, then also $\s \pre \t'$,
for all $\s \in T$ and $\t,\t' \in \Tmc$.

\LP
For every $\t \in \Tmc$ we define the unfolding
\[ \t \as = \{ \s \in T \mid \exists \t' \in \Tmc\. \t \unfolds \t'
   \mbox{\ and\ } \s \pre \t' \} \]

\begin{proposition}
$\t \as \in \Tinf$.
\end{proposition}

\proof
We have to show that $\t\as$ is an ideal.
It is non-empty, $\void\in \t\as$, and downward closed.
It is also directed: Let $\s,\r \in \t\as$.
Then there is $\t'$ with $\t \unfolds \t'$, $\s \pre \t'$
and $\t''$ with $\t \unfolds \t''$, $\r \pre \t''$.
As $\unfold$ is confluent, there is $\psi$ with $\t' \unfolds \psi$
and $\t'' \unfolds \psi$.
It follows $\s \pre \psi$ and $\r \pre \psi$,
therefore $\s \pj \r \pre \psi$ and $\s \pj \r \in \t\as$.
\eproof

\DEFI
We define an equivalence relation $\approx$ on types by:
$\s \approx \t$ iff $\s\as = \t\as$.
\EDEFI

\LP
$\approx$ is decidable \cite{Amadio/Cardelli}.

\subsection{Terms}

For every type $\t \in \Tmc$ 
there is a denumerable set $V^{\t}$ of variables of
type $\t$. The sets $V^{\t}$ are mutually disjoint. Their members are
denoted by $x^{\t}, y^{\t}, \ldots$ There is a set \CA{C} of constants
with types ${\it ctype}: \CA{C} \fu \Tmc$.

General untyped terms are built from variables and constants by application
$MN$ and ($\l$-)abstraction $\l x.M$, without regarding the types.
$\Lambda$ is the set of all untyped terms.

\LP
We give rules for the formation of typed terms; $M: \s$ means:
$M$ has type $\s$, $\s \in \Tmc$ :
\begin{deflisteng}{(const)}
\item[(const)] $c : {\it ctype} (c)$ for $c \in \CA{C}$
\item[(var)]   $x^{\s} : \s$
\item[(\fu I)] $M : \t \Ra \l x^{\s}.M : \s \f \t$
\item[(\fu E)] $M : \s \fu \t, N : \s \Ra MN : \t$
\item[($\approx$)] $M : \s, \s \approx \t \Ra M : \t$
\end{deflisteng}
Terms are considered equal modulo $\alpha$-conversion.
We abbreviate $\l x.\l y.M$ as $\l xy.M$.  Often type superscripts
of variables will be omitted. \CA{T} is the set of all typed terms.
The type of a typed term is unique up to $\approx$,
so the inference rules
could be given for type trees instead of types.
$\T_\s$ is the set of all terms with type $\s \in \Tmc$
or with type tree $\s \in \Tinf$.
$\T^c_\s$ is the corresponding set of all closed terms.
In the following chapters terms will always be understood to be typed.

For every type $\s$ we can define a fixed point combinator:
\[ Y_{\s} = \l y^{\s \f \s}.(\l x^{\m t.(t \f \s)}.y(x x))
                             (\l x^{\m t.(t \f \s)}.y(x x)) 
                              : (\s \f \s) \f \s \]

Remark: We have given a type system with rule ($\approx$) instead of
explicit conversion operators between the types $\m t.\s$ and
$\s[\m t.\s / t]$, called rep/abs, unfold/fold or elim/intro in
\cite{Winskel:book,Cosmadakis,Amadio/Cardelli,Gunter}. There are untyped
terms that can be typed in our system, but not in a system with explicit
conversion, even with the introduction of arbitrary rep/abs in the term.
E.g.\ let $M = Y(\l f x.f)$ and  $N = Y(\l f x y.f)$
in $(vM,vN)$. In this term, $M$ and $N$ must have the same type, which is
impossible in an abs/rep-system. In our system the types of
$M : \m t.\s \fu t$ and $N : \m t.\s \fu \s \fu t$
are equivalent.
Moreover our type system with rule $(\approx)$ has principle type schemes.
A system with the weaker congruence $\sim$, as the smallest congruence
(\wrt type constructors) such that $\m t.\s \sim \s[\m t.\s / t]$,
lacks this property \cite{Cardone/Coppo,Cardone/CoppoRec}.

\LP
Our special set of constants consists of the following symbols for
all types $\s,\t,\r$:
\begin{deflisteng}{${\pr pcase}_{\s,\t,\r}$}
\item[${\pr 0}_{\s,\t}:$] $\s \fu (\s + \t) $, also called ``inleft''
  in the literature
\item[${\pr 1}_{\s,\t}:$] $\t \fu (\s + \t) $, also called ``inright''
\item[${\pr case}_{\s,\t,\r}:$] $(\s + \t)\fu(\s\fu\r)\fu(\t\fu\r)\fu\r$,
                                sequential conditional
\item[${\pr pcase}_{\s,\t,\r}:$] $(\s + \t)\fu\r\fu\r\fu\r$,
     parallel conditional. Note the type different from \case's type.
\item[${\pr pair}_{\s,\t}:$] $\s\fu\t\fu(\s \times\t)$,
     $\pair\: x\: y$ is also written $(x,y)$
\item[${\pr fst}_{\s,\t}:$] $(\s \times \t)\fu\s$
\item[${\pr snd}_{\s,\t}:$] $(\s \times \t)\fu\t$
\item[$\Omega_{\s}:$] $\s$, the canonical undefined term of type $\s$.
     $\Omega_{\s}$ has the same denotational semantics as $Y_{\s}(\l x^{\s}.x)$.
     There are no reduction rules for $\O$.
\end{deflisteng}

We will frequently omit the type subscripts of the constants. The term rewriting
system will treat them as single symbols. Notice that we do not introduce
these operators by special term formation rules for the types 
$\s + \t$ and $\s \times \t$, as it is often done, but as constants of higher
order types that can be applied by normal application.
\0, \1, \pair\ are the constructors for building up the canonical terms
of type  $\s +\t$, $\s \times \t$ respectively.
{\pr case, pcase, fst, snd} are the corresponding evaluators.
We will usually write $\0$ instead of $\0\O$ and $\1$ instead of $\1\O$.

We could also include in our calculus separated sum types with a different
number of components than two. A special case would be the type constructor
\lift\ with just one type argument. It adds a new bottom element to the
domain of the type. The constants for this type constructor would be
$\ell_{\s}: \s \f (\lift\: \s)$ and ${\pr lcase}_{\s,\t}:
(\lift\: \s) \f (\s \f \t) \f \t$, corresponding to \0\ and \case.
We omit this type constructor as it can be treated analogously to +.

\LP
Examples of common types and their canonical terms:
\LP
$\void \approx \m t.t$ has just one element, denoted by $\Omega_{\pr void}$.
\LP
\begin{picture}(75,30)(-60,-10)
\put(-20,10){\mb[br]{
${\pr bool} \= \void + \void$}}
\put(0,0){\mb[b]{$\O$}}
\put(-10,10){\mb[b]{$\0\O$}}
\put(10,10){\mb[b]{$\1\O$}}
\dl (2,4)(10,9)
\dl (-2,4) (-10,9)
\end{picture}
\LP
\begin{picture}(110,30)(-80,0)
\put(-40,20){\mb[br]{
${\pr bitstream} \= \m t.t+t$}}
\put(0,0){\mb[b]{$\O$}}
\put(-20,10){\mb[b]{$\0\O$}} \put(20,10){\mb[b]{$\1\O$}}
\put(-30,20){\mb[b]{$\0(\0\O)$}}  \put(-10,20){\mb[b]{$\0(\1\O)$}}
\put(10,20){\mb[b]{$\1(\0\O)$}}   \put(30,20){\mb[b]{$\1(\1\O)$}}
\dl(2,4)(18,9)
\dl(19,14)(10,19)
\dl(21,14)(30,19)
\dl(9,25)(5,28)
\dl(11,25)(15,28)
\dl(29,25)(25,28)
\dl(31,25)(35,28)
\dl(-2,4)(-18,9)
\dl(-19,14)(-10,19)
\dl(-21,14)(-30,19)
\dl(-9,25)(-5,28)
\dl(-11,25)(-15,28)
\dl(-29,25)(-25,28)
\dl(-31,25)(-35,28)
\end{picture}
\LP
\begin{picture}(120,45)(-90,0)
\put(-40,30){\mb[br]{
${\pr nat} \= \m t.\void + t$,}}
\put(-40,25){\mb[br]{
the lazy natural numbers:}}
\put(0,0){\mb[b]{$\O$}}
\put(-10,10){\mb[br]{$0\cong\0\O$}} \put(10,10){\mb[b]{$\1\O$}}
\put(0,20){\mb[br]{$\mbox{succ}\: 0 \cong \1(\0\O)$}}
\put(20,20){\mb[b]{$\1(\1\O)$}}
\put(10,30){\mb[br]{$\mbox{succ}(\mbox{succ}\: 0) \cong \1(\1(\0\O))$}}
\put(30,30){\mb[b]{$\1(\1(\1\O))$}}
\dl (2,4)(10,9)
\dl (-2,4) (-10,9)
\dl(6,12)(-5,19)
\dl(13,14)(18,19)
\dl(14,23)(5,29)
\dl(24,24)(28,29)
\dl(30,35)(25,39)
\dl(34,35)(39,39)
\end{picture}
\LP
${\pr boollist} \= \m t.\void + ({\pr bool} \times t)$\\
{\pr boollist} is the type of lists of elements of {\pr bool},\\
\eg $\0_{{\pr void},{\pr bool}\times{\pr boollist}} \GO_{\pr void}:
    {\pr boollist}$, simply written as \0 without type subscripts
and undefined term \GO, the empty list,\\
\eg $\1_{{\pr void},{\pr bool}\times{\pr boollist}}
    (\1_{{\pr void},{\pr void}} \GO_{\pr void},
     \0_{{\pr void},{\pr bool}\times{\pr boollist}} \GO_{\pr void}):
    {\pr boollist}$, simply written as \1(\1,\0),
the list of one element $\1$.

Note that ``infinitely branching'' domains, like the flat domain of
natural numbers of PCF, cannot be defined in our type system because
the type constructor of coalesced sums is missing.

\subsection{Reduction}

We define a reduction relation $\f$ on terms. It performs a one-step
reduction of a single redex in any context. It is the least relation
satisfying:
\begin{deflisteng}{(\pcond 00)}
\item[($\beta$)] the $\beta$-reduction rule:\\
$(\l x.M)N \f M[x \subst N]$ for any terms $M$,$N$ and variable $x$,
where $M[x \subst N]$ is the substitution of $N$ for the free occurrences
of $x$ in $M$, with appropriate renaming of bound variables of $M$,

\item[ ] three context rules:

\item[(app)] $M \f M' \Longrightarrow MN \f M' N$,\\
             $N \f N' \Longrightarrow MN \f M N'$,

\item[($\l$)] $M \f M' \Longrightarrow \l x.M \f \l x.M'$,

\item[ ] and a set of applicative term rewriting rules for the constants,
where the variables $x,y,z,w$ denote arbitrary terms:

\end {deflisteng}

\[ \begin{array}{llcl}
(\cond \0) & \cond\:(\0 x)\:y\:z &\f& y\: x\\
(\cond \1) & \cond\:(\1 x)\:y\:z &\f& z\: x\\
(\pair 1) & \fst\:(\pair\: x \: y) &\f& x\\
(\pair 2) & \snd\:(\pair\: x \: y) &\f& y\\
(\pcond \0)& \pcond\:(\0 x) \: y \: z &\f& y\\
(\pcond \1)& \pcond\:(\1 x)\:y\:z &\f& z\\
(\pcond \0\0)& \pcond_{\s,\t,\r_0 + \r_1}\:x\:(\0 y)\:(\0 z) &\f&
               \0\:(\pcond_{\s,\t,\r_0}\:x\:y\:z)\\
(\pcond \1\1)& \pcond_{\s,\t,\r_0 + \r_1}\:x\:(\1 y)\:(\1 z) &\f&
               \1\:(\pcond_{\s,\t,\r_1}\:x\:y\:z)\\
(\pcond\! \times\!\times)& \pcond_{\s,\t,\r_1 \times \r_2}\:x\:(y_1, y_2)\:
                       (z_1, z_2) &\f& (\pcond_{\s,\t,\r_1}\:x\:y_1\:z_1,
                                        \pcond_{\s,\t,\r_2}\:x\:y_2\:z_2)\\
(\pcond \f)& (\pcond_{\s,\t,\r_1\f\r_2}\:x\:y\:z)\:w &\f&
             \pcond_{\s,\t,\r_2}\:x\:(y\:w)\:(z\:w)
\end{array} \]
$\rast$ is the reflexive, transitive closure of $\f$.

Note   the order of parameters of \cond: $y$ is the \0-part,
$z$ is the \1-part.
The functionality of \cond\ permits the definition of the usual
evaluators ``outleft'' and ``outright'', so that we need not introduce them
with reduction rules:
\[ \begin{array}{lcl}  \label{out}
{\pr out0}_{\s,\t} &:& (\s + \t) \f \s\\
{\pr out0} &\=& \l x.\cond\:x\:(\l y.y)\:\O\\
{\pr out1}_{\s,\t} &:& (\s + \t) \f \t\\
{\pr out1} &\=& \l x.\cond\:x\:\O\:(\l y.y)
\end{array} \]

\pcond\ is not a sequential function, as it forces its three arguments to be
reduced in parallel. As soon as the ``boolean value'' of its first
argument appears, a reduction with rule (\pcond 0) or (\pcond 1) can be made.
As soon as the second and the third argument convey the same piece of
information, namely a constructor \0, \1\ or \pair, this piece of information
can be drawn out of the \pcond-expression according to rule (\pcond \0\0),
(\pcond \1\1) or (\pcond $\times \times$). If the second and the third
argument are of functional type, then the argument $w$ of the \pcond-expression
can be drawn in according to rule (\pcond $\f$), so that $(y\:w)$ and
$(z\:w)$ can deliver constructor information before the evaluation of
$x$ is finished.
Note that \pcase\ appears on the right sides of its rules
(\pcase\0\0)--(\pcase$\f$).
It performs a recursion on the type tree of its second and third argument.
We could think of a parallel conditional with the same type as \case.
But for such a conditional it is more difficult to implement this
recursion by rewrite rules; in fact we would need conditioned rewrite
rules with $\l$-abstractions and \outz, \outi\ in the right sides.

\begin{proposition}
Our reduction relation $\f$ fulfills the subject reduction property:
If $M: \s$ and $M \rast N$, then also $N:\s$.
\end{proposition}

\proof
The property can be checked for each reduction rule.

\begin{theorem}[Confluence]
$\f$ is confluent (Church-Rosser) on typed terms:\\
For any typed term $M \in \CA{T}$ with $N \last M \rast P$ there is a
term $Q$ with $N \rast Q \last P$. ($N,P,Q$ are also typed with equivalent
types due to the subject reduction property.)
\end{theorem}

Note that the restriction of $M$ to typed terms is essential, as can be seen
with the term $\pcond\:x\:(\0y)\:(\0 z)\:w$. This term is not typable,
as $(\0 y)$ is not of function type. It reduces to
$\pcond\:x\:(\0 y\:w)\:(\0 z\:w)$ by rule (\pcond $\f$), and to
$\0 (\pcond\:x\:y\:z)\:w$ by rule (\pcond\0\0). This critical pair does not
converge to a common reduct.

\proof 
We will use the confluence theorem of \cite{Mueller:lconfluence}:
For every left-linear, not variable-applying ATRS (applicative term
rewriting system) with reduction relation $\f$ and every $\f$-closed
set $T$ of terms: If $\f$ is confluent on the applicative terms of $T$
then $\f$ is confluent on $T$. We explain the notions of this theorem
in our context:

The applicative terms are the terms without any $\l$-abstraction, \ie
they are built only from variables, constants and application.
An ATRS is a set of pairs $\LR$ of applicative terms, where $L$ is no
variable and all variables of $R$ appear in $L$, too. In our case, the ATRS
is the set of reduction rules (\case\0) \ldots (\pcase$\f$).
Together with $\beta$-reduction and the context rules (app) and ($\l$)
it determines the reduction relation $\f$ on terms of $\Lambda$.
It is left-linear, \ie every variable has at most one occurrence in each
left side of the rules. It is not variable-applying, \ie no left side of
any rule contains a subterm of the form $(x M)$, where $x$ is a variable.
In our case, $T$ will be the set \CA{T} of typed terms. \CA{T} is 
$\f$-closed, \ie for every $M \in \CA{T}$ the
following hold: \\
1) $M \f M' \Rightarrow M' \in \CA{T}$, the subject reduction property,\\
2) every subterm of $M$ is in \CA{T},\\
3) for every occurrence $u$ of an abstraction in $M$, $M/u = \l \ldots$,
there is a variable $x$ not occurring in $M$ with
$M[u \lf x] \in \CA{T}$.\\
We use the same notations for occurrences of subterms and replacement
at an occurrence as \cite{Huet,Mueller:lconfluence}.
In condition 3 we chose a new variable of the appropriate type.

Now it remains to prove the confluence of $\f$ on the set \CA{A} of
applicative terms of \CA{T}, \ie the confluence of the ATRS alone,
without $\beta$-reduction. Our theorem, the confluence of $\f$ on all
terms of \CA{T}, follows by the cited theorem.

From now on, $\f$ is the reduction relation on applicative terms of $\Lambda$.
We will first prove that $\f$ is locally confluent on \CA{A} via
convergence of critical pairs, then prove that $\f$ is noetherian
(terminating, strongly normalizing) and conclude the confluence of
$\f$ on \CA{A} by Newman's Lemma (Lemma 2.4 of \cite{Huet}).
{\de Local (or weak) confluence} of $\f$ on a set $T$ of terms means:
For any $M \in T$ with $N \lf M \f P$ there is a term $Q$ with
$N \rast Q \last P$.

Notice that the sufficient conditions for confluence in \cite{Huet}
that check only convergence of critical pairs, without termination,
are not applicable here:
Huet's Lemma 3.3 is almost applicable (Corollary: Any left-linear
parallel closed term rewriting system is confluent), but it demands
of the critical pair:\\
$y\:w \lf (\pcond\:(\0 x)\:y\:z)\:w \f \pcond\:(\0 x)\:(y\:w)\:(z\:w)$
that there should be a parallel reduction step:
$y\:w \f \pcond\:(\0 x)\:(y\:w)\:(z\:w)$.
Note that the right term of a critical pair is defined by a reduction
{\em at the root}. The lemma demands a parallel reduction step from
the left to the right term, not an arbitrary reduction. But in our
example there is only a reduction in the opposite direction.
\cite[Corollary 3.2]{Toyama} gives a sufficient condition more general
than Huet's Lemma 3.3; it is also not applicable here by the same reason.

For the proof of local confluence of $\f$ on \CA{A} we will apply a
generalized version of Lemma 3.1 of \cite{Huet}:
``For any term rewriting system \CA{R}: The relation $\fR$ is locally
confluent iff for every critical pair $(P,Q)$ of \CA{R} we have
$P \downarrow Q$, \ie $P$ and $Q$ have a common reduct.''
This lemma cannot be applied directly, as 
the non-typable, non-convergent critical pair
given before this proof
shows us. It should state local confluence on certain subsets of
terms which resemble sets of well-typed terms, similar to the $\f$-closed
sets of terms above. This leads us to:

\DEFI
A subset $T$ of terms is called $\fR${\em -complete} for a term rewriting
system with reduction relation $\fR$ if for every $M \in T$ the
following hold:
\begin{deflisteng}{1)}
\item[1)] $M \fR M' \Rightarrow M' \in T$,
\item[2)] every subterm of $M$ is in $T$,
\item[3)] for every set of occurrences $u_1,\ldots,u_n$ of the same
subterm $N$ in $M$, \ie $M/u_i = N$ for all $i$, there is a variable $x$
not occurring in M with $M[u_1 \lf x] \ldots [u_n \lf x] \in T$.
\end{deflisteng}
\EDEFI

Let us recall the definition of critical pairs of a term rewriting system.
\DEFI
Let $\ST$,$\LR$ be two rules whose variables are renamed such that
$L$ and $S$ have disjoint variable sets. Let $u$ be an occurrence in $L$
such that $L/u$ is no variable and $L/u$ and $S$ are unifiable with
substitution $\m$ as the most general unifier.
The superposition of $\ST$ on $\LR$ in $u$ determines the
{\em critical pair} $(P,Q)$ defined by
$P = (\m L)[u \lf \m T]$, $Q = \m R$.
It is $P \lf \m L \f Q$. We call $\m L$ an {\em overlap} of the critical
pair $(P,Q)$.
\EDEFI
Our generalization of Huet's Lemma 3.1 is now:
\begin{lemma}
For any term rewriting system \CA{R} and $\fR$-complete subset $T$ of terms:
The reduction relation $\fR$ is locally confluent on $T$ iff for
every critical pair $(P,Q)$ of \CA{R} with an overlap in $T$ we have
$P \downarrow Q$.
\end{lemma}

\proof(sketch)
The proof is essentially the proof of Lemma 3.1 in \cite{Huet}.
The ``only if'' part is trivial again. For the ``if'' part we add the
assumption $M \in T$. Case 1 (disjoint redexes) and case 2a (prefix 
redexes that do not overlap) are the same as in \cite{Huet}.
Case 2b deals with overlapping redexes: An overlap of the critical pair
is obtained from the subterm $M/u_1$ by replacing some subterms by variables.
It is $M/u_1 \in T$ according to condition 2 of $\fR$-completeness.
The replacement of subterms by variables is possible according to condition 3
of $\fR$-completeness, so that the overlap is in $T$.
Thus $P \downarrow Q$ by hypothesis, and the proof proceeds as in \cite{Huet}.
\eproof

We use the lemma to show local confluence of $\f$ on \CA{A}.
\CA{A} is $\f$-complete. Eight critical pairs with an overlap in \CA{A}
remain to be checked for convergence:

\[ \begin{array}{rcl}
(\0 y) \lf & \pcond\:(\0 x)\:(\0 y)\:(\0 z) & \f \0 (\pcond\:(\0 x)\:y\:z)\\
(\1 y) \lf & \pcond\:(\0 x)\:(\1 y)\:(\1 z) & \f \1 (\pcond\:(\0 x)\:y\:z)\\
(\0 z) \lf & \pcond\:(\1 x)\:(\0 y)\:(\0 z) & \f \0 (\pcond\:(\1 x)\:y\:z)\\
(\1 z) \lf & \pcond\:(\1 x)\:(\1 y)\:(\1 z) & \f \1 (\pcond\:(\1 x)\:y\:z)\\
(y_1,y_2) \lf & \pcond\:(\0 x)\:(y_1,y_2)\:(z_1,z_2) & 
                \f (\pcond\:(\0 x)\:y_1\:z_1, \pcond\:(\0 x)\:y_2\:z_2)\\
(z_1,z_2) \lf & \pcond\:(\1 x)\:(y_1,y_2)\:(z_1,z_2) & 
                \f (\pcond\:(\1 x)\:y_1\:z_1, \pcond\:(\1 x)\:y_2\:z_2)\\
y\:w \lf & \pcond\:(\0 x)\:y\:z\:w & \f \pcond\:(\0 x)\:(y\:w)\:(z\:w)\\
z\:w \lf & \pcond\:(\1 x)\:y\:z\:w & \f \pcond\:(\1 x)\:(y\:w)\:(z\:w)
\end{array} \]

We prove now that $\f$ is noetherian on applicative terms.
(This will also be used in the proof of Lemma 5.3.)
We define a mapping $\ph$ from applicative terms to $\{2,3,\ldots\}$
inductively by the following equations:

\[ \begin{array}{lcl}
\ph M &=& 2, \mbox{\ if $M$ is a variable or a constant}\\
\ph (\0 M) &=& 2 \cdot \ph M\\
\ph (\1 M) &=& 2 \cdot \ph M\\
\ph (\pcond\: M) &=& 2 \cdot \ph M\\
\ph (\pcond\:MN) &=& 2 \cdot \ph M \cdot \ph N\\
\ph (\pcond\:MNP)&=& 2 \cdot \ph M \cdot \ph N \cdot \ph P\\
\ph (\pair\:M) &=& 2 + \ph M\\
\ph (\pair\:MN) &=& 2 + \ph M + \ph N\\
\ph (MN) &=& (\ph M)^{\ph N}, \mbox{\ for all other applications}\; MN
\end{array} \]

By simple computations we show for every reduction rule $\LR$ that
$\ph L > \ph R$, where variables of the rule stand for arbitrary terms.
The two interesting rules are:
\LP 
$ (\pcond\times\times)\;\; \pcond\:x\:(\pair\:y_1\:y_2)\:(\pair\:z_1\:z_2)
   \;\f\; \pair\:(\pcond\:x\:y_1\:z_1)\:(\pcond\:x\:y_2\:z_2)$
\LP
$ \ph L = 2 \cdot \ph x \cdot (2+\ph y_1+\ph y_2)\cdot(2+\ph z_1+\ph z_2)$
\LP
$ \ph R = 2+2\cdot\ph x\cdot\ph y_1\cdot\ph z_1 +
          2\cdot\ph x\cdot\ph y_2\cdot\ph z_2 $
\LP
$ (\pcond\f)\;\; (\pcond\:x\:y\:z)\:w \;\f\; \pcond\:x\:(y\:w)\:(z\:w)$
\LP
$ \ph L = 2^{\ph w}\cdot (\ph x)^{\ph w}\cdot (\ph y)^{\ph w}\cdot 
(\ph z)^{\ph w}$
\LP
$ \ph R = 2 \cdot \ph x \cdot \ph(y\:w) \cdot \ph(z\:w) $

For the last rule (and some other) we need the fact that
$(\ph M)^{\ph N} \geq \ph(MN)$ for all terms $M,N$,
which we prove by a case analysis over the term $M$.

It remains to show that a reduction at any position decreases the
$\ph$-value of a term. We prove that
\[ \ph N > \ph N' \Rightarrow  \ph (MN) > \ph (MN')\]
and that
\[ \ph M > \ph M' \mbox{\ and\ } M \f M' \Rightarrow \ph(MN) > \ph(M'N) \]
for all terms $M,N,M',N'$ by a case analysis over $M$.

We have now proved that $M \f N \Rightarrow \ph M > \ph N$.
Thus there are no infinite reduction chains. From this and the local
confluence of $\f$ on \CA{A} follows by Newman's Lemma the
confluence of $\f$ on \CA{A}.
As explained above, the confluence of $\f$ on all typed terms follows
from the theorem of \cite{Mueller:lconfluence}.
\eproof

\section{Prime systems}

We introduce prime systems as concrete representations of domains,
together with operations on them corresponding to the type constructors
$+,\p,\f$. The results of this chapter are taken from \cite{Larsen/Winskel}
, where they were given for the more general information systems.
\DEFI
A {\de prime system} $\ps{A} = (A,\con,\lep)$ consists of\\
a set $A$
(the {\de primes}, denoted by $a,b,c$),\\
a reflexive and symmetric binary relation $\con$ on $A$
(the {\de consistency}),\\
and a partial order $\lep$ on $A$ (the {\de entailment}),\\
such that for all $a,b,c \in A$: If $a \con b$ and $c \lep b$, then
$a \con c$.\\
\PSys\ is the class of all prime systems.
\EDEFI

Prime systems were first introduced in \cite{NielsenPW} under the name
``event structures'', where the elements of $A$ were interpreted as events
of a computation process. (Instead of consistency there was the dual
conflict relation.) Here we chose a different name because we do not
interpret the elements of $A$ as events, but as pieces of information,
as in information systems.
A prime is an elementary, indivisible piece of information about
data elements. The relation
$a \lep b$ means that whenever $b$ is valid of an element,
then so is $a$. 
$a \con b$ means that both primes $a$ and $b$ may be valid of an element.

Every prime system determines an information system in the sense of
\cite{Larsen/Winskel}: 
The set of tokens is $A$.
A finite subset $X$ of $A$ is consistent ($X\in \mbox{Con}$)
iff for all $a,b \in X$, $a\con b$.
For $X \in \mbox{Con}$ and $a \in A$ we define $X \vdash a$
iff $\exists b \in X\. a \lep b$.
We use the simpler prime systems instead of information systems as they
are just suited for our data types.

\DEFI
The {\de elements} of a prime system $\ps{A} = (A,\con,\lep)$
are the subsets $d \sub A$ that are downward closed:
$a \lep b \e b \in d \Ra a \in d$,
and consistent: $a \con b$ for all $a,b \in d$.

$\eleps{A}$ is the set of elements of $\ps{A}$.
We call $\eleps{A}$, ordered by $\sub$, the {\de domain} of $\ps{A}$.
The least element $\emptyset$ is also denoted by $\bot$.

For $X \sub A$ we write 
$\downc{X}\: = \{a\in A \st \exists b \in X\. a \lep b\}$,
also $\downc{a}$ for $\downc{\{a\}}$.
The {\de finite elements} of $\ps{A}$ are defined as the elements of the form
$\downc{X}$ for finite $X \sub A$.
\EDEFI

We will give the characterization of the domains of prime systems from
\cite{NielsenPW}.
First some domain theoretic definitions.

\DEFI
Let $(D,\suq)$ be a partial order. A subset of $D$ is
{\de pairwise consistent} iff any two of its elements have an upper bound
in $D$.
$(D,\suq)$ is {\de coherent} iff every pairwise consistent subset of $D$
has a lub.

$p \in D$ is a {\de complete prime} iff for every $S \sub D$, if the
lub $\bigsqcup S$ exists and $p \suq \bigsqcup S$, then there is
$d \in S$ with $p \suq d$.\\
$(D,\suq)$ is {\de prime algebraic} iff for every $d \in D$ the set
$\{ p \suq d \st p \mbox{\ is a complete prime} \}$
has $d$ as its lub.
\EDEFI

\begin{theorem} \cite{NielsenPW}
Let $\ps{A} = (A,\con,\lep)$ be a prime system.
Then $(\eleps{A},\sub)$ is a prime algebraic coherent partial order.
Its complete primes are the elements $\downc{a}$ for $a \in A$.\\
It follows that $(\eleps{A},\sub)$ is also an algebraic cpo.
Its isolated (or finite, compact) elements are the finite elements
defined above.

Conversely, let $(D,\suq)$ be a prime algebraic coherent partial order.
Let $P$ be the set of complete primes of $D$, and $a \con b$ iff
$a,b \in P$ have an upper bound.
Then $\ps{P}=(P,\con,\suq)$ is a prime system
with $(\ele{\ps{P}},\sub)$ isomorphic to $(D,\suq)$.
\end{theorem}

This theorem explains our name for ``primes''.
From this characterization we only need the fact that the domain of
a prime system is a cpo, \ie has lubs of directed subsets. These lubs
are the set unions of the elements.

As in \cite{Larsen/Winskel} we define a complete partial order on the
class of prime systems and continuous operations on prime systems.

\DEFI
Let $\ps{A}=(A,\con_A,\lep_A)$ and $\ps{B}=(B,\con_B,\lep_B)$
be prime systems.
We define $\ps{A} \sust\ps{B}$ iff $A\sub B$ and for all $a,b \in A$:
$a \con_A b \Lra a \con_B b$ and $a \lep_A b  \Lra a \lep_B b$.
\EDEFI

$\ps{A} \sust \ps{B}$ means that $\ps{A}$ is a {\em subsystem} of $\ps{B}$:
$A\sub B$ and $\con_A, \lep_A$ are the restrictions of $\con_B, \lep_B$
on $A$.
If $\ps{A} \sust \ps{B}$ and $A=B$, then $\ps{A}=\ps{B}$.

\begin{theorem}
$\sust$ is a partial order with $\ubot = (\emptyset,\emptyset,\emptyset)$
as least element.
If $\Achain$ is an $\omega$-chain of prime systems
$\Ai$, then 
\[ \cupAi = (\bigcup_i A_i, \bigcup_i \con_i, \bigcup_i \lep_i) \]
is the lub of the chain.
\end{theorem}

\proof
Clearly $\sust$ is a partial order, $\ubot$ is the least element.\\
Now for the chain $\ps{A}_i$ let
$\ps{A} = (A, \con,\lep) = 
 (\bigcup_i A_i, \bigcup_i \con_i, \bigcup_i \lep_i) $ .

$\ps{A}$ is an upper bound of the chain:
$A_i \sub A$ for all $i$. Let $a,b \in A_i$.
If $a \con_i b$, then $a \con b$.
Conversely, if $a \con b$, then $a,b \in A_j$ and $a \con_j b$ for some $j$.
If $j \leq i$, then $\ps{A}_j \sust \ps{A}_i$;
if $i \leq j$, then $\ps{A}_i \sust \ps{A}_j$.
In either case follows $a \con_i b$.
Analogously we show $a \lep_i b \Lra a\lep b$.

$\ps{A}$ is the least upper bound of the chain:
Let $\ps{B} = (B,\con_B,\lep_B)$ be an upper bound of the chain.
Then $A= \bigcup_i A_i \sub B$.
Let $a,b \in A$. Then $a,b \in A_i$ for some $i$.
We have $a \con b \Lra a \con_i b \Lra a\con_B b$ and
$a \lep b \Lra a\lep_i b \Lra a \lep_B b$.
\eproof

\LP
We extend $\sust$ to n-tuples of prime systems.

\DEFI
For $n\geq 1$, $\PSys^n$ are all n-tuples $(\ps{A}_1,\ldots,\ps{A}_n)$
of prime systems.
We define 
\[ (\psA_1,\ldots,\psA_n) \sust (\psB_1,\ldots,\psB_n) \Lra
\psA_1 \sust \psB_1 \e \ldots \e \psA_n \sust \psB_n. \].
\EDEFI
\vspace{-8mm}

\begin{prop}
$\sust$ is a partial order on $\PSys^n$ with $(\ubot,\ldots,\ubot)$
as least element.
All increasing $\omega$-chains in $(\PSys^n,\sust)$ have a least upper bound
taken coordinate-wise.
\end{prop}

\DEFI
Let $F: \PSys^n \f \PSys$ be an operation on prime systems.\\
$F$ is called {\de monotonic} iff
$\psA \sust \psB \Ra F(\psA) \sust F(\psB)$ for all $\psA,\psB \in \PSys^n$.\\
$F$ is called {\de continuous} iff it is monotonic and for any $\omega$-chain
of prime systems
$\Achain$ in $\PSys^n$, $F(\cupAi) = \bigcup_i F(\psA_i)$.
(Since $F$ is monotonic, $F(\psA_i), i\geq 0,$ is an ascending chain
and $\bigcup_i F(\psA_i)$ exists.)
\EDEFI

\begin{prop}
$F: \PSys^n \f \PSys$ is monotonic (continuous) iff it is 
monotonic (continuous) in each argument separately (\ie considered as a
function in any of its arguments, holding the others fixed).
\end{prop}

Thus to show that an operation is monotonic or continuous we have to show
that some unary operations are monotonic or continuous.
The following lemma will help in these proofs.

\begin{defi}
$F: \PSys \f \PSys$ is {\de continuous on prime sets} iff for any 
$\omega$-chain of prime systems $\Achain$ each prime of $F(\cupAi)$ is a prime of
$\bigcup_i F(\psA_i)$.
\end{defi}

\begin{lemma}\label{contin}
$F: \PSys \f \PSys$ is continuous iff $F$ is monotonic and
continuous on prime sets.
\end{lemma}

\proof
The ``only if'' part is obvious.\\
``if'': Let $\Achain$ be an $\omega$-chain of prime systems.
From $\psA_i \sust \cupAi$ and monotonicity follows
$F(\psA_i) \sust F(\cupAi)$.
Then $\bigcup_i F(\psA_i) \sust F(\cupAi)$.
As $F$ is continuous on prime sets, the primes of $\bigcup_i F(\psA_i)$
are the same as those of $F(\cupAi)$.
Therefore they are the same prime systems.
\eproof

\subsection*{Operations on prime systems}

We give continuous operations on prime systems corresponding to our
syntactic type constructors $\void, +, \p,\f$.\\
Corresponding to \void\ is the prime system 
$\ubot=(\emptyset,\emptyset,\emptyset)$.
It has the only element $\emptyset = \bot$.

\subsection* {Separated sum $+$}

\begin{defi}
Let $\psA_0 = (A_0,\con_0,\lep_0)$ and
 $\psA_1 = (A_1,\con_1,\lep_1)$ be prime systems.
Define $\psA_0 + \psA_1 = (B,\con,\lep)$ by
\EQN
B &=& B_0 \cup B_1 \\
& &\mbox{\ where\ } B_0=\{0\}\cup(\{0\}\cp A_0)
                   \mbox{\ and\ }   B_1=\{1\}\cup(\{1\}\cp A_1),\\
a\con b &\Lra& (a,b \in B_0 \mbox{\ and if\ }a=(0,a_0),b=(0,b_0), \mbox{\ then\ }
               a_0 \con_0 b_0)\\
&\mbox{or}& (a,b \in B_1 \mbox{\ and if\ }a=(1,a_1),b=(1,b_1),\mbox{\ then\ }
               a_1 \con_1 b_1),\\
a\lep b &\Lra& a=0,\; b\in B_0 \\
&\mbox{or}& a=1,\; b\in B_1\\
&\mbox{or}& a=(0,a_0),\; b=(0,b_0),\; a_0 \lep_0 b_0\\
&\mbox{or}& a=(1,a_1),\; b=(1,b_1),\; a_1 \lep_1 b_1 .
\EEQN
\end{defi}

\begin{prop}
$\psA_0 + \psA_1$ is a prime system.
Its domain is
\[ \ele{\psA_0 + \psA_1} = \{\emptyset\}
\cup \{ \{0\} \cup (\{0\} \cp d) \st d \in \ele{\psA_0} \}
\cup \{ \{1\} \cup (\{1\} \cp d) \st d \in \ele{\psA_1} \}. \]
\end{prop}
We abbreviate the element $\set{0}$ as $0$ and $\set{1}$ as $1$.

\begin{theorem}
$+$ is continuous on $(\PSys,\sust)$.
\end{theorem}

\proof
It is easy to show that + is continuous in its first and second
argument, using
Lemma \ref{contin}.
\eproof

\subsection*{Product $\p$}

\begin{defi}
Let $\psA_0 = (A_0,\con_0,\lep_0)$ and
 $\psA_1 = (A_1,\con_1,\lep_1)$ be prime systems.
Define $\psA_0 \p \psA_1 = (B,\con,\lep)$ by
\EQN
B &=& (\set{0}\cp A_0) \cup (\set{1} \cp A_1),\\
a \con b &\Lra& a=(0,a_0),\;b=(0,b_0),\; a_0 \con_0 b_0\\
&\mbox{or}    & a=(1,a_1),\;b=(1,b_1),\; a_1 \con_1 b_1\\
&\mbox{or}&     a=(0,a_0),\;b=(1,b_1)\\
&\mbox{or}&     a=(1,a_1),\;b=(0,b_0),\\
a \lep b &\Lra& a=(0,a_0),\;b=(0,b_0),\;a_0 \lep_0 b_0\\
&\mbox{or}&     a=(1,a_1),\;b=(1,b_1),\;a_1 \lep_1 b_1 .
\EEQN
\end{defi}

\begin{prop}
$\psA_0 \p \psA_1$ is a prime system. Its domain is
\[ \ele{\psA_0 \p \psA_1} = 
   \set{(\set{0}\cp d)\cup(\set{1}\cp e) \st d\in \ele{\psA_0}
    \e e\in \ele{\psA_1}} \]
\end{prop}

\begin{theorem}
$\p$ is continuous on $(\PSys,\sust)$.
\end{theorem}

\proof
It is easy to show that $\p$ is continuous in its first and second
argument, using
Lemma \ref{contin}.
\eproof

\subsection*{Function space $\f$}

\begin{defi}
Let $\psA=(A,\con_A,\lep_A)$ and $\psB=(B,\con_B,\lep_B)$
be prime systems. (We leave out the indexes in the following.)\\
We define $\psA \f \psB = (C,\con,\lep)$:
\\ $C=\ov{A} \cp B$, where $\ovA$ is the set of all finite subsets
of $A$ that are pairwise consistent and incomparable,
$\ovA = \set{X \sub A \st X \mbox{\ finite} \et
 \forall a,b\in X\. a\con b \e (a\lep b \Ra a=b)}$.\\
Let $(X,a),(Y,b)\in C$.
\[(X,a)\con (Y,b) \Lra (X\con Y \Ra a\con b),\]  
where $X\con Y \Lra \forall a\in X, b\in Y\. a\con b$.
\[(X,a)\lep (Y,b) \Lra Y\lep X \et a\lep b,\]
where $Y\lep X \Lra Y \sub \downc{X}$, \ie 
$\forall a\in Y\. \exists b\in X\. a\lep b$.
\end{defi}

\begin{prop}
$\psA \f \psB$ is a prime system.
\end{prop}
 
\proof\\
$\con$ is reflexive and symmetric. $\lep$ is reflexive.
\LP
$\lep$ is antisymmetric:\\
Let $(X,a)\lep(Y,b)$ and $(Y,b)\lep(X,a)$. We show $(X,a)=(Y,b)$.\\
We have $a\lep b$ and $b\lep a$, so $a=b$.\\
From $X\lep Y$ and $Y\lep X$ we conclude $X\sub Y$:
\\ Let $x\in X$. 
There is $y\in Y$ with $x\lep y$, and $x'\in X$ with $y\lep x'$.
So $x\lep x'$, and $x=x'$ by the condition on $X$. Hence $x=y\in Y$.\\
Similarly we conclude $Y\sub X$.
\LP
$\lep$ is transitive:\\
Let $(X,a)\lep (Y,b)\lep(Z,c)$. We show $(X,a)\lep(Z,c)$.\\
We have $a\lep b\lep c$, so $a\lep c$.
From  $Z\lep Y\lep X$ we conclude $Z\lep X$:\\
Let $z\in Z$. There is $y\in Y$ with $z\lep y$,
and $x\in X$ with $y\lep x$.

\LP
It remains to show:
If $(X,a)\con(Y,b)$ and $(Z,c)\lep (Y,b)$, then $(X,a)\con (Z,c)$.\\
Suppose $X\con Z$. Then $X\con Y$:
Let $x\in X, y\in Y$. $Y\lep Z$, therefore $\exists z\in Z\. y\lep z$.
It is $x\con z$, hence $x\con y$.\\
We get $a\con b$ and $c\lep b$, therefore $a\con c$.
\eproof

The elements of $\psA\f\psB$ correspond to
the continuous functions from domain $\elepsA$ to $\elepsB$.

\begin{prop}
Let $r\in \ele{\psA \f \psB}$.
Then $\fl{r}: \elepsA \f \elepsB$ given by
\[ \fl{r}(d) = \set{a\st \exists X \sub d\. (X,a)\in r} \mbox{\ for\ }
   d \in \elepsA \]
is a continuous function from the domain $\elepsA$ to $\elepsB$.
\end{prop}

\proof
We show $\fl{r}(d)\in \elepsB$.\\
$\fl{r}(d)$ is consistent:
Let $a,b\in \fl{r}(d)$. There is $X\sub d$ with $(X,a)\in r$
and $Y\sub d$ with $(Y,b)\in r$. As $(X,a)\con(Y,b)$ and
$X\con Y$, we conclude $a\con b$.\\
$\fl{r}(d)$ is downward closed:
Let $b\in \fl{r}(d)$ and $a\lep b$.
There is $Y\sub d$ with $(Y,b)\in r$.
It is $(Y,a)\lep(Y,b)$, so $(Y,a)\in r$ and $a\in \flr(d)$.

\LP
$\flr$ is monotonic, obviously.\\
$\flr$ is continuous: Let $D$ be a directed subset of $\elepsA$.
 \begin{eqnarray*}
\bigcup_{d\in D} \flr(d) &=&
   \set{a \st \exists d\in D\. \exists X\sub d\. (X,a)\in r}\\
&=& \set{a \st \exists X \sub \bigcup D\. (X,a)\in r},
    \mbox{\ because the $X$ are finite}\\
&=& \flr(\bigcup D)
\end{eqnarray*} 
\eproof

For cpos $(D,\sub)$ and $(E,\sub)$, let $([D \f E],\sub)$ be the
cpo of continuous functions from $D$ to $E$, ordered pointwise by $\sub$.
We will also write $f: D \f E$ for $f\in[D \f E]$, and
$f: D\f E\f F$ for $f\in [D\f [E \f F]]$.
For $f: D\f E$ and $d\in D$ we will
usually    write $f\:d$ instead of $f(d)$, as in the syntax of
the lambda calculus. Here also application is associated to the
left, \ie $f\:d\:e = (f\:d)\:e$.
We will frequently write $r\:d$ instead of $\flr(d)$.
It  is clear from  the context that
the function between domains is meant.

\begin{prop}
Let $f: \elepsA \f \elepsB$ be monotonic and $A$ be the set of primes of
$\psA$.
Then the {\de prime set of} $f$,
\[  \Pr(f) = \set{(X,a) \st X\in \ovA \e a \in f(\downc{X})}, \]
is an element of $\ele{\psA \f \psB}$.
\end{prop}

\proof\\
$\Pr(f)$ is consistent:
Let $(X,a),(Y,b) \in \Pr(f)$ and assume $X\con Y$.
Then $\downc{(X \cup Y)} \in \elepsA$.
As $a \in f(\downc{X})$ and $b \in f(\downc{Y})$,
we have $a,b \in f(\downc{(X \cup Y)})$, by monotonicity of $f$.
Therefore $a\con b$.

\LP
$\Pr(f)$ is downward closed:
Let $(X,a)$ and $(Y,b)$ be primes of $\psA \f \psB$,
$(Y,b)\in \Pr(f)$ and $(X,a) \lep (Y,b)$.
From $Y\lep X$ follows $\downc{Y} \sub \downc{X}$.
Then $b \in f(\downc{X})$, as $b\in f(\downc{Y})$ and $f$ is monotonic.
As $a\lep b$, also $a\in f(\downc{X})$
and $(X,a)\in \Pr(f)$.
\eproof

\begin{theorem}
For all prime systems $\psA,\psB$ the map
\[\ele{.}: (\ele{\psA \f \psB}, \sub) \f ([\elepsA \f \elepsB], \sub)\]
is an isomorphism of cpos. The map \Pr\ is its inverse.\\
Therefore the complete primes and isolated elements of $[\elepsA \f \elepsB]$
are the images under $\ele{.}$ of the corresponding elements of
$\ele{\psA \f \psB}$.
\end{theorem}

\proof
We show that for all $r\in \ele{\psA \f \psB}$, $\Pr(\flr) =r$:
 \begin{eqnarray*}
(X,a)\in \Pr(\flr) &\Lra& a\in \flr(\downc{X})\\
&\Lra& \exists Y\sub \downc{X}\. (Y,a)\in r\\
&\Lra& (X,a)\in r, \mbox{\ because $(X,a)\lep(Y,a)$ and $r$ is
downward closed}
\end{eqnarray*} 

\LP
We show that for all $f\in [\eleps{A} \f \elepsB]$, $\fl{\Pr(f)}=f$:\\
Let $A,B$ be the set of primes of $\psA$ and   $\psB$, resp.
Let $d\in \elepsA$ and $a\in B$.
\begin{eqnarray*}
a \in \fl{\Pr(f)} (d) &\Lra&
   \exists X\sub d\. X\in \ovA \e (X,a)\in \Pr(f)\\
&\Lra& \exists X \sub d\. X\in \ovA \e a \in f(\downc{X})\\
&\Lra& a\in f(d)
\end{eqnarray*} 
We prove the last equivalence:\\
$\Ra$: $\downc{X} \sub d$ and $f$ is monotonic.\\
$\La$: Let $D= \set{\downc{Y} \st Y \mbox{\ finite}\et Y\sub d}$.
$D$ is a directed set in $\elepsA$. $\bigcup D = d$.
Since $f$ is continuous, there is some finite $Y$ with $Y\sub d$
and $a\in f(\downc{Y})$.
Let $X$ be the set of maximal primes of $Y$.
We get $X\sub d$, $X \in \ovA$, $\downc{Y}=\downc{X}$
and $a\in f(\downc{X})$.

So the map $\ele{.}$ is one-to-one, \Pr\ is its inverse.
It remains to show that $\ele{.}$ and \Pr\ respect the partial order $\sub$:
\[ \mbox{For all\ } r,s\in \ele{\psA \f \psB}:
   r\sub s \Lra \forall d\in \elepsA\. \flr(d) \sub \fl{s}(d) \]
$\Ra$ is obvious.\\
$\La$: Let $(X,a)\in r$. Then $a\in \flr(\downc{X})$.
As $a\in \fl{s}(\downc{X})$, there is $Y\sub \downc{X}$ with $(Y,a)\in s$.
As $Y \lep X$, also $(X,a)\in s$.
\eproof

\begin{theorem}
$\f$ is continuous on $(\PSys,\sust)$.
\end{theorem}

\proof\\
1) $\f$ is monotonic in its first argument:\\
Let $\psA_0 = (A_0,\con_0,\lep_0) \sust \psA_0'=(A_0',\con_0',\lep_0')$,
$\psA_1 = (A_1,\con_1,\lep_1)$ be prime systems and
$\psA_0 \f \psA_1 = (B,\con,\lep)$, $\psA_0' \f \psA_1 = (B',\con',\lep')$.
\\ We have to prove: $\psA_0 \f \psA_1 \sust \psA_0' \f \psA_1$.\\
First we show: $B= \ov{A_0} \cp A_1 \sub \ov{A_0'} \cp A_1 = B'$.
\\ Let $X\in \ov{A_0}$.
For all $a,b\in X$: $a\con_0' b$ and $(a\lep_0' b \Ra a=b)$.
Therefore $X\in \ov{A_0'}$.

\LP
Now let $(X,a),(Y,b) \in B$.
\begin{eqnarray*}
(X,a)\con (Y,b) &\Lra& (X\con_0 Y \Ra a\con_1 b)\\
                &\Lra& (X\con_0' Y \Ra a\con_1 b)\\
                &\Lra& (X,a) \con' (Y,b)\\
(X,a) \lep(Y,b) &\Lra& Y \lep_0 X \et a\lep_1 b\\
                &\Lra& Y \lep_0' X \et a\lep_1 b\\
                &\Lra& (X,a) \lep' (Y,b)
\end{eqnarray*} 

\LP
2) $\f$ is continuous on prime sets in its first argument:\\
Let $\Achain$ be an $\omega$-chain of prime systems with $\Ai$,
and $\psB$ be a prime system.\\
Let $(X,b)$ be a prime of $(\cupAi) \f \psB$.
Then $X\in \ov{\bigcup_i A_i}$.
Since $X$ is finite, $X\sub A_n$ for some $n$.
For all $a,c \in X$, $a\con_n c$ and $(a \lep_n c \Ra a=c)$,
because $\psA_n \sust \cupAi$.
So $X\in \ov{A_n}$ and $(X,b)$ is a prime of $\bigcup_i (\psA_i \f \psB)$.

\LP
3) $\f$ is monotonic in its second argument:\\
Let $\psA_0 = (A_0,\con_0,\lep_0)$,
$\psA_1 = (A_1,\con_1,\lep_1) \sust \psA_1'=(A_1',\con_1',\lep_1')$
be prime systems and
$\psA_0 \f \psA_1 = (B,\con,\lep)$,
$\psA_0 \f \psA_1' = (B',\con',\lep')$.
We have to show: $\psA_0 \f \psA_1 \sust \psA_0 \f \psA_1'$.\\
$B = \ov{A_0} \cp A_1 \sub \ov{A_0} \cp A_1' = B'$.\\
Now let $(X,a),(Y,b) \in B$.
\begin{eqnarray*}
(X,a)\con (Y,b) &\Lra& (X\con_0 Y \Ra a\con_1 b)\\
                &\Lra& (X\con_0 Y \Ra a\con_1' b)\\
                &\Lra& (X,a) \con' (Y,b)\\
(X,a) \lep(Y,b) &\Lra& Y \lep_0 X \et a\lep_1 b\\
                &\Lra& Y \lep_0 X \et a\lep_1' b\\
                &\Lra& (X,a) \lep' (Y,b)
\end{eqnarray*} 

\LP
4) $\f$ is continuous on prime sets in its second argument:\\
Let $\Achain$ be an $\omega$-chain of prime systems with $\Ai$,
and $\psB=(B,\con,\lep)$ be a prime system.\\
The set of primes of $\psB \f (\cupAi)$ is
$\ov{B} \cp (\bigcup_i A_i) = \bigcup_i(\ov{B} \cp A_i)$,
the set of primes of $\bigcup_i (\psB \f \psA_i)$.
\eproof
               
\section{Denotational semantics}

\subsection{Semantics of types}

We give a semantic interpretation of the type trees of $\Tinf$
as prime systems. 
So we do not solve recursive domain equations directly,
but define the semantics of a recursive type $\t\in \Tmc$
by the semantics of its unfolding $\t\as$.

\begin{defi}
The sequence of maps $\P_n : \Tinf \f \PSys$, $n\geq 0$,
is defined inductively by:
\ARR{lcl}
\P_0(\s) &=& \ubot \mbox{\ for all\ } \s \in \Tinf,\\
\P_{n+1}(\void) &=& \ubot,\\
\P_{n+1}(\s \tc \t) &=& \P_n(\s) \tc \P_n(\t) \mbox{\ for\ }
   \tc\: \in \set{+,\p,\f} \mbox{\ and\ } \s, \t \in \Tinf.
\EARR
Define $P_i(\s)$ as the prime set of $\P_i(\s)$.
\end{defi}

\begin{prop}
For all $\s\in \Tinf$, $n\geq 0$: $\P_n(\s) \sust \P_{n+1}(\s)$.
\\ (This proposition depends only on the monotonicity of the
operations $+,\p,\f$ on prime systems.)
\end{prop}

\proof by induction on n. Trivial for $n=0$.\\
Now assume that for some $n\geq 0$:
$\forall \s\in \Tinf\. \P_n(\s) \sust \P_{n+1}(\s)$.\\
We prove $\P_{n+1}(\s) \sust \P_{n+2}(\s)$ for all cases of $\s$:\\
$\P_{n+1}(\void) = \ubot \sust \P_{n+2}(\void)$.\\
$\P_{n+1}(\s \tc \t) = \P_n(\s) \tc \P_n(\t) \sust
  \P_{n+1}(\s) \tc \P_{n+1}(\t) = \P_{n+2}(\s \tc \t)$
for $\tc\: \in \set{+,\p,\f}$.
\eproof

\medskip\LP
This permits to give the semantics of type trees:

\begin{defi}
Define the map $\P: \Tinf \f \PSys$ by 
$\P(\s) = \bigcup_i \P_i(\s)$.\\
$P(\s)$ is the set of primes of $\P(\s)$.
\end{defi}

\begin{prop}
\mbox{  }\\[-7mm]
\ARR{lcl}
\P(\void) &=& \ubot \\
\P(\s \tc \t) &=& \P(\s) \tc \P(\t) \mbox{\ for\ }
  \tc\: \in \set{+,\p,\f} \mbox{\ and\ } \s,\t\in \Tinf
\EARR
(This proposition depends on the continuity of the operations
$+,\p,\f$ on prime systems.)
\end{prop}

\proof  Clearly $\P(\void) = \ubot$.
\EQN
\P(\s \tc \t) &=& \bigcup_i (\P_{i+1}(\s \tc \t))\\
&=& \bigcup_i (\P_i(\s) \tc \P_i(\t))\\
&=& (\bigcup_i \P_i(\s)) \tc (\bigcup_i \P_i(\t))\\
&=& \P(\s) \tc \P(\t).
\EEQN
\eproof

\DEFI
The domain for a type tree $\s\in\Tinf$ is $D_\s=\ele{\P(\s)}$,\\
the domain for a type  $\s\in\Tmc$ is $D_\s =\ele{\P(\s\as)}$.\\
For $d\in D_\s$, $\s \in \Tinf$, we define the $n$-th {\de projection}
of $d$ as $d|_n = d \cap P_n(\s)$.
\EDEFI

Note that the primes of $\P(\s)$ are expressions of finite size
and therefore structural induction may be applied to them.
More precisely: For a prime $a\in P(\s)$ let $\level(a)$
be the least $i$ such that $a\in P_i(\s)$.\\
If $(0,a)\in P(\s+\t)$, then $a\in P(\s)$ and $\level(a)<\level(0,a)$.\\
If $(1,a)\in P(\s+\t)$, then $a\in P(\t)$ and $\level(a)<\level(1,a)$.\\
The same holds for $\s\p\t$ instead of $\s+\t$.\\
If $(X,a)\in P(\s\f\t)$, then for all $x\in X$: $x\in P(\s)$ and
$\level(x) < \level(X,a)$,
and $a\in P(\t)$ and $\level(a) < \level(X,a)$.\\
Therefore definitions and proofs for primes may be given by induction on
their parts with smaller level.

\subsection{Semantics of terms}

We will define the semantics function $\Sem$ for terms.
As usual we need environments:
Let $V=\bigcup_{\t\in\Tmc} V^\t$ be the set of all term variables of any type.
An {\de environment} is a function $\ep: V \f \bigcup_{\s\in\Tmc} D_\s$
such that $\ep(x^\s)\in D_\s$ for all $x^\s \in V$.
$\Env$ is the set of all environments.
It is a cpo under the pointwise order $\sub$. 
Its least element is denoted by $\bot$, $\bot(x)=\bot$ for all $x$.
For any environment $\ep$, $\ep[x\mto d]$ is the environment
$\ep'$ with $\ep'(x)=d$ and $\ep'(y)=\ep(y)$ for $y\neq x$.

For every constant $c$ we will give a continuous function on domains.
This function is then transformed by \Pr\ into an element of the
prime system corresponding to the type of $c$.
We need versions of \Pr\ for functions with 2 and 3 arguments:

Let $f: \elepsA \f (\elepsB \f \eleps{C})$ be continuous for
prime systems $\psA,\psB,\ps{C}$.
Define $\Pr_2(f) \in \ele{\psA \f (\psB \f \ps{C})}$
by $\Pr_2(f) = \Pr(\Pr \circ f)$,
where $(f \circ g) x = f(g(x))$.
Note that $\Pr \circ f$ is continuous since \Pr\ is continuous
as an order isomorphism.
It is $(\Pr_2(f))\,a\,b = \fl{(\fl{\Pr_2(f)}\,a)}\,b = f\,a\,b$.

Let $f: \elepsA \f(\elepsB \f (\eleps{C} \f \eleps{D}))$ be continuous
for prime systems $\psA,\psB,\ps{C},\ps{D}$.
Define $\Pr_3(f) \in \ele{\psA \f (\psB \f (\ps{C}\f \ps{D}))}$
by $\Pr_3(f) = \Pr(\Pr_2 \circ f)$.
Note that $\Pr_2 \circ f$ is continuous as $\Pr_2$ is continuous.
It is $(\Pr_3(f))\,a\,b\,c =
\fl{(\fl{(\fl{\Pr_3(f)}\,a)}\,b)}\,c = f\,a\,b\,c$.

\DEFI
We define the semantics function $\Sem : \Term    \f (\Env\f
\bigcup_{\s\in\Tmc} D_\s)$
by structural induction on the term argument.
We write $\S{M}$ and $\Se{M}$, for $M\in\Term$, $\ep\in\Env$.
It is $\S{M} \in [\Env \f D_\s]$ for $M:\s$, see the following proposition.

\[ \begin{array}{lclcl}
\Se{\0_{\s,\t}} &=& \Pr(0), &\mbox{with}& 0: D_\s \f D_{\s+\t}\\
&&&&  0d = \set{0}\cup(\set{0}\cp d)\\
\Se{\1_{\s,\t}} &=& \Pr(1), &\mbox{with}& 1: D_\t \f D_{\s+\t}\\
&&&&  1d = \set{1}\cup(\set{1}\cp d)\\
\Se{\case_{\s,\t\r}} &=& \Pr_3(\scase), &\mbox{with}&
   \scase: D_{\s+\t} \f D_{\s\f\r} \f D_{\t\f\r} \f D_\r \\
&&&&  \scase\:d\,f\,g = \left\{
         \begin{array}{ll}
         \bot, & \mbox{if\ } d=\bot\\
         \fl{f}e, & \mbox{if\ } d= 0e\\
         \fl{g}e, & \mbox{if\ } d =1e
         \end{array}
         \right.     \\     
\Se{\pcase_{\s,\t,\r}} &=& \Pr_3(\spcase), &\mbox{with}&
   \spcase: D_{\s+\t} \f D_\r \f D_\r \f D_\r \\
&&&&  \spcase\:a\,b\,c = \left\{
         \begin{array}{ll}
         b \cap c, & \mbox{if\ } a = \bot\\
         b, & \mbox{if\ } a=0a'\\
         c, & \mbox{if\ } a=1a'
         \end{array}
         \right.     \\
\Se{\pair_{\s,\t}} &=& \Pr_2(\spair),  &\mbox{with}&
    \spair: D_\s \f D_\t \f D_{\s\p\t} \\
&&&&  \spair\:d\,e = (\set{0}\cp d) \cup (\set{1}\cp e) \\
\Se{\fst_{\s,\t}} &=& \Pr(\sfst), &\mbox{with}& \sfst: D_{\s\p\t} \f D_\s \\
&&&&  \sfst\,(\spair\:d\,e) = d \\
\Se{\snd_{\s,\t}} &=& \Pr(\ssnd), &\mbox{with}& \ssnd: D_{\s\p\t} \f D_\t \\
&&&&  \ssnd\,(\spair\:d\,e) = e 
\end{array} \]

\[ \begin{array}{lclcl}
\Se{\O_\s} &=& \bot & & \\
\Se{x}  &=& \ep(x) & & \\
\Se{\l x^\s.M} &=&
  \multicolumn{3}{l}{\Pr(d\in D_\s \mto \S{M}(\ep[x \mto d])),} \\
&&\multicolumn{3}{l}{\mbox{where ($d\in D \mto$ exp) denotes the function}} \\
&&\multicolumn{3}{l}{\mbox{that maps each $d\in D$ to exp}} \\
\Se{MN} &=& \multicolumn{3}{l}{\fl{\Se{M}}\: (\Se{N})}
\end{array} \]
\EDEFI

\PROP
For all terms $M:\psi$, $\S{M} \in [\Env \f D_\psi]$.
\EPROP
 \proof by structural induction on $M$.\\
$\bullet$ Let $M$ be a constant:\\
It is easy to check that the given function on domains
is  continuous and that the semantics of $M$ is in the
appropriate domain. We show this only for $M=\pcase_{\s,\t\r}$:

\spcase\ is monotonic (and continuous) in its first argument,
since $b\cap c\sub b$ and $b\cap c \sub c$.
\spcase\ is continuous in its second (third) argument: This is clear
for the cases $a=0a'$ and $a=1a'$.
In the case $a=\bot$ it follows from the continuity of $\cap$.
Now $\spcase: D_{\s+\t} \f D_\r \f D_\r \f D_\r$ is continuous, therefore
\EQN
\Se{\pcase_{\s,\t,\r}} = \Pr_3(\spcase) &\in&
   \ele{\P((\s+\t)\as) \f \P(\r\as) \f \P(\r\as) \f \P(\r\as)} \\
&=&   D_{(\s+\t)\f\r\f\r\f\r} . 
\EEQN
If $\pcase_{\s,\t,\r}: \psi$, then $\psi \approx (\s+\t)\f\r\f\r$,
and $\S{\pcase_{\s,\t,\r}} \in [\Env \f D_\psi]$.

\medskip
\LP
$\bullet$ Let $M=x^\s$:\\
$\S{x^\s} = (\ep \mto \ep(x^\s)): \Env \f D_\s$ is continuous.

\medskip
\LP
$\bullet$ Let $M= \l x^\s.N: \s \f \t$:\\
Then $N:\t$, and $\S{N} \in [\Env \f D_\t]$ follows by induction hypothesis.
Let $\ep \in \Env$ and $f = (d\in D_\s \mto \S{N}(\ep[x\mto d]))$.
$f$ is continuous, because $\ep[x\mto .]$ and $\S{N}$ are continuous.
So $f\in [D_\s\f D_\t]$, and
\[ \Se{\l x.N} = \Pr(f) \in \ele{\P(\s\as) \f \P(\t\as)} = D_{\s\f\t} .\]
It remains to show that $\S{\l x.N}$ is continuous.\\
It is monotonic: Let $\ep,\ep' \in \Env$ and $\ep \sub\ep'$. Then
\EQN
\Se{\l x.N} &=& \Pr(d\in D_\s \mto \S{N}(\ep[x\mto d])) \\
 &\sub& \Pr(d\in D_\s \mto \S{N}(\ep'[x\mto d])),
   \mbox{\ as $\S{N}$ and \Pr\ are monotonic} \\
&=& \S{\l x.N}\ep'
\EEQN
Let $E$ be a directed set of environments.
\EQN
\S{\l x.N}(\bigcup_{\ep\in E} \ep) &=&
    \Pr(d\in D_\s \mto \S{N}((\bigcup_{\ep\in E} \ep)[x \mto d])) \\
&=& \Pr(d\in D_\s \mto \S{N}(\bigcup_{\ep\in E} (\ep[x \mto d]))) \\
&=& \Pr(d\in D_\s \mto \bigcup_{\ep\in E} \S{N}(\ep[x \mto d])),
    \mbox{\ as $\S{N}$ is continuous} \\
&=& \bigcup_{\ep\in E} \Pr(d\in D_\s \mto \S{N}(\ep[x\mto d])),
    \mbox{\ as \Pr\ is continuous} \\
&=& \bigcup_{\ep\in E} \Se{\l x.N}
\EEQN
\LP
$\bullet$ Let $M=NP$, $N: \s \f\t$, $P: \s$:
\\ By induction hypothesis we have $\S{N} \in [\Env \f D_{\s\f\t}]$
and $\S{P} \in [\Env\f D_\s]$.
Let $\ep\in \Env$.
Then $\fl{\Se{N}}\in D_\s\f D_\t$ and $\Se{P}\in D_\s$,
hence $\Se{NP} \in D_\t$.
$\S{NP}$ is continuous because $\S{N}$, $\S{P}$ and $\fl{.}$ are
continuous.
So we get $\S{NP} \in [\Env \f D_\t]$.
\eproof

\subsection{Soundness of the semantics}

We show that reduction does not change the semantics of terms.
First we prove the Substitution Lemma.

\LEMMA \label{subst} (Substitution Lemma)
\[ \Se{M[x \subst N]} = \S{M} (\ep[x \mto \Se{N}]), \]
for all appropriately typed terms $M,N$, and all $\ep\in \Env$.
\ELEMMA

\proof by induction on the structure of $M$,
see Lemma 2.12 of \cite{Gunter}.
\eproof

\THEOREM
[Soundness]
If $M,N \in \Term$ and $M \rast N$, then $\S{M} = \S{N}$.
\ETHEOREM

\proof
It is clear that the semantics of a term is not changed by replacing
a subterm by a term with the same semantics.
We have the properties:
\ARR{lcl}
\S{M}= \S{M'} &\Ra& \S{MN}=\S{M'N} \\
\S{N}= \S{N'} &\Ra& \S{MN}=\S{MN'} \\
\S{M}= \S{M'} &\Ra& \S{\l x.M}=\S{\l x.M'}
\EARR
So if $\S{M}=\S{M'}$, then $\S{C[M]} = \S{C[M']}$ for any context $C[\:]$.\\
It can be easily checked that each reduction rule does not change the semantics.
For the $\beta$-rule this follows from the Substitution Lemma.
\eproof
\section{Approximation Theorem}

For every term $M$ we will define a set $\A{M}$ of normal forms
that approximate the reducts of $M$.
$\A{M}$ can be seen as the syntactic value of $M$ or the
B\"ohm tree of $M$.
We will prove the Approximation Theorem:
$\Se{M} = \bigcup_{A\in \A{M}} \Se{A}$.
Thus the semantics of $M$ is entirely determined by the
normal form approximations of $M$.

There are three methods in the literature to prove the
Approximation Theorem:
\cite[Th. 3.1.12]{Berry} proves it for PCF
and \cite{Wadsworth} for the untyped lambda calculus,
both with the aid of a labelled $\l$-calculus.
\cite{Mosses/Plotkin} proves it for the untyped $\l$-calculus
by two other methods: by an intermediate semantics
and by inclusive predicates.
We will give an inclusive predicate proof,
modified for the recursively typed $\l$-calculus and prime systems.

\LP
First we use the constant $\O$ to define the usual
$\O$-prefix partial order on terms:
\DEFI
For every $\s\in \Tinf$, $\pre$ is the least relation on $\T_\s$ satisfying:\\
$\O\pre M$ for every $M\in \T_\s$,\\
$x\pre x$ for every variable or constant $x$,
\\ $M\pre M' \Ra \l x.M \pre \l x.M'$,\\
$M\pre M' \e N\pre N' \Ra MN \pre M'N'$.\\
If $M,N\in \T_\s$ have an upper bound under $\pre$, then
$M \pj N$ is defined as their least upper bound.
\EDEFI
It is clearly: $M\pre N \Ra \S{M} \sub \S{N}$.

\DEFI
Let $\s\in\Tinf$.
$\N_{\s}$ is the set of normal form terms of $\T_\s$.
Normal forms are denoted by $A,B,\ldots$.\\
Let $A\in \N_\s$, $M\in \T_\s$.\\
$A$ is a {\de direct approximation} of $M$, $A\dap M$,
iff $\forall N\.(M\rast N \Ra A\pre N)$.\\
$A$ is an {\de approximation} of $M$, $A\ap M$,
iff $\exists N\. M\rast N \et A\dap N$.\\
$\A{M}$ denotes the set of approximations of $M$.\\
We abbreviate $\Sae{M} = \bigcup_{A\apind M} \Se{A}$.
\EDEFI

A direct approximation of $M$ conveys a fixed syntactic information
about $M$: It is in normal form and is part of all reducts of $M$.
If $A\dap M$ and $M\rast N$, then $A\dap N$.
We want to show that $\A{M}$ is an ideal.
Therefore we need the following lemma, which relies on the fact
that all applicative terms have a normal form.

\LEMMA
If $A\dap M$ and $B\dap M$, then $A\pj B$ exists and is a normal form,
and $A\pj B \dap M$.
\ELEMMA
\proof
$A\pj B$ exists because $A\pre M$ and $B\pre M$.
Now assume that $A\pj B$ is not a normal form.
Then there is an occurrence $u$ in $A\pj B$ such that $(A\pj B)/u$
is a redex.

First assume that it is a $\beta$-redex:
$(A\pj B)/u$ is of the form $(\l x.N) P$.
Then either $A/u$ is of the form $(\l x.N') P'$, or $B/u$ is of this form.
This contradicts the assumption that $A$ and $B$ are normal forms.

Now assume that $(A\pj B)/u$ is a redex of a constant, corresponding
to one of the rules (\case 0) -- (\pcase $\f$).
Let $L=M/u$.
Let $u_i$, $1\leq i\leq n$, be a sequence of all the outermost
occurrences of $\l$-abstractions in $L$.
Let $x_i$, $1\leq i\leq n$, be a sequence of distinct variables
that do not occur in $L$. (The type of $x_i$ should be that of $L/u_i$.)
Let $K= L[u_1\lf x_1,\ldots,u_n\lf x_n]$.
$K$ is an applicative term, \ie it does not contain any $\l$-abstraction.
As $\f$ is strongly normalizing (noetherian) on applicative terms,
there is a normal form $K'$ of $K$, $K\rast K'$.
It is $L=K[x_1\subst (L/u_1),\ldots,x_n\subst (L/u_n)]$, 
the result of the replacement of the $x_i$ by $L/u_i$.
Let   $L'=K'[x_1\subst (L/u_1),\ldots,x_n\subst (L/u_n)]$.
Then $L\rast L'$.
As $K'$ is a normal form and the $L/u_i$ are $\l$-abstractions,
$L'$ is not a redex of a constant.\\
It is $M\rast M[u\lf L']$, as $L\rast L'$.
As $A\dap M$ and $B\dap M$, we have $A\pj B \pre M[u\lf L']$.
Therefore $(A\pj B)/u \pre L'$.
This contradicts the fact that $L'$ is not a redex of a constant.

So in every case we deduced a contradiction from the assumption that
$A\pj B$ is not a normal form.
Clearly $A\pj B \dap M$.
\eproof

\THEOREM
$\A{M}$ is an ideal under $\pre$, \ie it is non-empty,
downward closed and directed.
\ETHEOREM
\proof
We have $\O\in \A{M}$.\\
$\A{M}$ is downward closed: If $A\ap M$ and $B\pre A$, then $B\ap M$.\\
$\A{M}$ is directed: Let $A \ap M$ and $A' \ap M$.
There is $N$ with $M\rast N \e A\dap N$,
and $N'$ with $M\rast N' \e A' \dap N'$.
By confluence there is a term $P$ with $N\rast P$ and $N'\rast P$.
Then $A\dap P$ and $A'\dap P$.
By the preceding lemma, $A\pj A'$ is a normal form and $A\pj A' \dap P$.
Hence $A\pj A' \ap M$.
\eproof
\LP
With this proposition $\A{M}$ is an element of the ideal completion
of $\N_\s$ (under $\pre$); it can be seen as a B\"ohm tree of $M$.
\LP
Let us first discuss our definition of approximation and compare it
with different approaches in the literature:
\DL{3)}
\item[1)]
The treatment of PCF in \cite{Berry} is different:
The approximations are obtained by reducing only $\beta$- and
$Y$-redexes. The constants are treated like variables;
redexes of rules for constants are not reduced.
They are only interpreted semantically in the B\"ohm tree.
This approach is only possible because the reduction of constant redexes
can be postponed after the reduction of $\beta$- and $Y$-redexes.
In our case constants operate on higher order types as well,
therefore the reduction of constant redexes is intertwined
with $\beta$-reduction.

\item[2)]
$\A{M}$ is not minimal: In many cases there is a proper subset of
$\A{M}$ with the same semantics;
\eg for $M=\l x.\O$ or $M= \O N$ the approximation $\O$ is sufficient.
$\A{M}$ was defined to give ``all possible'' normal form information
about $M$.
The questions arise: In which sense is $\A{M}$ maximal?
[My conjecture is: For every directed set $S$ of minimum normal forms
of $M$ (def.\ below), if $S$ has the same semantics as $\A{M}$, then
$S \sub \A{M}$.]
Is a smaller set of approximations definable with the same semantics,
that gives a substantially stronger Approximation Theorem?

In the presence of parallel operations there is in general no least
approximation with the same semantics: Consider
\[ M = \l x.\pcase\:x\:(\case\:x\:\O\:(\l y.\1))\:\1 : \bool \f \bool .\]
$\Sb{M}$ is the function that maps $1\mto 1$, $0\mto\bot$.
Both $ \l x.\pcase\:\O\:(\case\:x\:\O\:(\l y.\1))\:\1$ and 
$\l x.\pcase\:x\:\O\:\1$ are minimal approximations of $M$ with
the same semantics as $M$.

\item[3)]
In the presence of \pcase\ it is not possible to define the approximations
by an analogue of head normal forms. We will make this statement precise
after the proof of the Approximation Theorem.
We will also give analogues of head normal forms for the sequential
calculus without \pcase.

\EDL

\LP
We now prove two useful lemmas about approximations.

\LEMMA \label{conv}
If $M\downarrow N$, then $\A{M}=\A{N}$ and $\Sae{M}=\Sae{N}$.
\ELEMMA
\proof
Let $M \rast P \last N$.
Assume $A\ap M$. Then there is $M'$ with $M\rast M'$ and $A \dap M'$.
By confluence there is $L$ with $M'\rast L \last P$.
Then $A\dap L$ and $A\ap N$.
This shows $\A{M} \sub \A{N}$.
Symmetrically $\A{M} \sup \A{N}$.
\eproof

\LEMMA \label{constant}
Let $c M_1\ldots M_n$ be a term where $c$ is a constant and there
are no reducts $M_i\rast M_i'$, $1\leq i \leq m \leq n$,
with $c M_1' \ldots M_m'$ a redex.
Then
\[ \Sae{c M_1\ldots M_n} = (\Sb{c})\:(\Sae{M_1})\dots (\Sae{M_n}).\]
\ELEMMA
\proof
\EQN
\Sae{c M_1\ldots M_n} &=& \bigcup \set{\Se{A} \st A\ap c M_1\ldots M_n}\\
&=& \bigcup \set{\Se{c A_1\ldots A_n} \st A_1 \ap M_1 \e \ldots
    \e A_n \ap M_n} \\
&=& (\Sb{c})\:(\Sae{M_1}) \ldots (\Sae{M_n})
\EEQN
We have used the fact that $A\ap c M_1\ldots M_n$
iff $A = c A_1 \ldots A_n$ with some $A_i \ap M_i$;
as no $c M_1 \ldots M_m$, $m \leq n$, can be reduced to a redex.
\eproof

\THEOREM[Approximation Theorem]
For all terms $M$ and environments $\ep$: \[\Se{M} = \Sae{M}.\]
\ETHEOREM
$\Sae{M}\sub \Se{M}$ follows from $\Se{A}\sub \Se{M}$ for $A\ap M$.
This is a consequence of soundness and of monotonicty of $\Sem$
\wrt $\pre$.
We want to prove the remaining inclusion
$\Se{M}\sub \Sae{M}$ by structural induction on $M$.
Therefore we use inclusive predicates (logical relations), also used in
\cite{Mosses/Plotkin} to prove the analogous theorem (limiting
completeness) for the untyped $\l$-calculus.
We define the inclusive predicates on the sets of primes
$P(\s)$ of the type interpretations $\P(\s)$:

\DEFI
For every $\s\in \Tinf$ and $\ep \in \Env$ we define a relation
$\ips\: \sub P(\s)\cp\T_\s$.
$a\ips M$ is defined by structural induction on $a$,
\ie in terms of propositions $a'\ipt M'$, where $a'$ is a part
of $a$ with smaller level.\\
There are the following cases for $\s$ and the primes:

\begin{tabbing}
$\s=\t\f\r:$ \= $(X,a)\ip{\t\f\r}M$ \= $\Lra$ \= \kill
$\s = \t+\r:$ \> $0\ip{\t+\r} M$ \> $\Lra 0\in \Sae{M} $\\[\sma]
\> $(0,a) \ip{\t+\r} M$ \> $\Lra (0,a) \in \Sae{M} \et a\ipt \Outz(M)$ \\[\sma]
\> $1\ip{\t+\r} M$ \> $\Lra 1\in \Sae{M} $\\[\sma]
\> $(1,a) \ip{\t+\r} M$ \> $\Lra (1,a) \in \Sae{M} \et a\ipr \Outi(M)$ \\[\sma]
\> where $\Outz(M)$ abbreviates the term $\case\:M\:(\l y.y)\:\O$,\\
\>  and $\Outi(M)$ the term $\case \:M\:\O\:(\l y.y)$.\\[\med]
$\s=\t\p\r:$ \> $(0,a)\ip{\t\p\r} M$
     \> $\Lra (0,a)\in \Sae{M} \et a\ipt \fst\:M$ \\[\sma]
\> $(1,a)\ip{\t\p\r} M$
     \> $\Lra (1,a)\in \Sae{M} \et a\ipr \snd\:M$ \\[\med]   
$\s=\t\f\r:$ \> $(X,a)\ip{\t\f\r} M$
   \> $\Lra (X,a) \in \Sae{M} \et$ \\[\sma]
\> \> \> $\forall N\in \T_\t\. (X\ipt N \Ra a \ipr MN)$
\end{tabbing}

For every set $X$ of primes $X\ipt N$ means:
$\forall b\in X\. b\ipt N$.
\EDEFI

Intuitively $a\ips M$ means that $a\in \Sae{M}$ and that the
relation is maintained in all contexts formed by \Outz, \Outi,
\fst, \snd\ and application on related arguments.
\LP
We have to prove a few lemmas for the Approximation Theorem.

\LEMMA \label{leq}
If $a\lep b \et b\ips M$, then also $a\ips M$.
\ELEMMA
\proof by structural induction on $b$.
In every case we have $a\in \Sae{M}$.
\LP
$\bullet$ $ \s=\t+\r:$\\
The case $a=0$, $b=(0,b')$ is clear.\\
Now let $a=(0,a')$, $b=(0,b')$.
Then $a'\lep b'$ and $b'\ipt \Outz(M)$.
By \ih follows $a'\ipt \Outz(M)$.\\
The cases $a=1, b=(1,b')$ and $a=(1,a'), b=(1,b')$ are analogous.
\LP
$\bullet$ $ \s=\t\p\r$ is like $\s=\t+\r$
\LP
$\bullet$ $ \s=\t\f\r:$\\
Let $a=(X,a')$, $b=(Y,b')$.
It is $Y\lep X$ and $a'\lep b'$.\\
For all $N\in \T_\t$ the following implications hold:
\EQN
X\ipt N &\Ra& Y\ipt N, \mbox{\ by \ih}\\
&\Ra& b' \ipr MN, \mbox{\ as\ } (Y,b')\ips M \\
&\Ra& a'\ipr MN, \mbox{\ by \ih}
\EEQN
Therefore $a=(X,a') \ips M$.
\eproof

\LEMMA \label{converge}
If $a\ips M$ and $M\downarrow N$, then also $a\ips N$.
\ELEMMA
\proof by structural induction on $a$.\\
We have $\Sae{M} = \Sae{N}$ by Lemma \ref{conv},
therefore $a\in \Sae{N}$.
\LP
\BU $\s =\t+\r:$
\\ Let $a=(0,a')$.
Then $a'\ipt \Outz(M)$.
By \ih follows $a' \ipt \Outz(N)$, so $a\ips N$.\\
$a=(1,a')$ is analogous.
\LP
\BU $\s=\t\p\r$ is like $\s=\t+\r$.
\LP
\BU $\s=\t\f\r:$\\
Let $a=(X,a')$. For all $P\in \T_\t$:
\EQN
X\ipt P &\Ra& a'\ipr MP, \mbox{\ as\ } a\ips M\\
&\Ra& a'\ipr NP, \mbox{\ by induction hyp., as\ } MP \downarrow NP
\EEQN
Therefore $a \ips N$.
\eproof
\medskip
\LP
We also need the new notion of passive term:
\DEFI
A term $M$ is a {\de redex part} iff
$M = \l x.N$ for some $x$ and $N$, or there is some typed
left-hand side $L$ of a rule (\case\0)\ldots (\pcase$\f$) and a subterm
$L'$ of $L$ such that $L'\neq L$, $L'$ is no variable
and $M$ is obtained from $L'$ by replacing variables by terms of the
same type.
\LP
This means: $M$ is a redex part iff $M$ is of one of the following 
forms:\\[\med]
$\l x.N,\; \0,\; \0 N,\; \1,\; \1 N,$\\
$\pair,\; \pair\:N_1,\; \pair\:N_1\:N_2,\; \fst,\; \snd,$\\
$\case,\; \case\:(\0 N),\; \case\:(\0 N_1)\:N_2,\;
          \case\:(\1 N),\; \case\:(\1 N_1)\:N_2,\;$\\
$\pcase,\; \pcase\:N_1,\; \pcase\:N_1\:(\0 N_2),\; \pcase\:N_1\:(\1 N_2),\;
  \pcase\:N_1\:(N_2, N_3),$\\
$\pcase\:N_1\:N_2$ with $N_2: \t\f\r$, 
  $\pcase\:N_1\:N_2\:N_3$ with $N_2,N_3: \t\f\r$.\\
(Note the type restrictions of the last two forms: They are parts of the
left-hand side of rule (\pcase$\f$).)

\LP
A term $M$ is called {\de passive} iff there is no redex part $N$
with $M\rast N$.
\EDEFI

No reduct of a passive term is able to interact with a context in the
reduction of a redex. 
Simple examples of passive terms are the variables.
The following two lemmas state the needed properties of passive terms.

\LEMMA \label{passive}
\mbox{ }
\DLE{3)}
\item[1)]
If $M$ is passive and $MN\rast P$, then $P=M'N'$ with 
$M\rast M'$ and $N\rast N'$.
\item[2)]
If $M$  is passive, then $MN$ is also passive for all $N$.
\item[3)]
If $M$ is passive, then 
$\Sae{MN} = \fl{\Sae{M}}\:(\Sae{N})$ for all $N$.
\EDLE
\ELEMMA
\proof
\DLE{3)}
\item[1)]
The proof is by induction on the length $n$ of the reduction $MN\rast P$.\\
It is clear for $n=0$.\\
Induction step: Let $MN\rast P\f Q$ be a reduction of length $n+1$.
By \ih $P=M'N'$ with $M\rast M'$ and $N\rast N'$.
$M'$ is no redex part.
Therefore either $Q=M''N'$ with $M'\f M''$ or $Q=M'N''$ with $N'\f N''$.

\item[2)]
Let $MN\rast P$.
By part 1) we have $P=M'N'$ with $M\rast M'$.
As $M'$ is not a redex part, $P$ is not a redex part either.
(There is no rule with a variable-applying left-hand side 
$x M_1\ldots M_n$.)

\item[3)]
For all $A$ we have:
\EQN
A \ap MN &\Lra& \exists P\. MN\rast P \e A\dap P \\
&\Lra& \exists M',N'\. M\rast M' \e N\rast N' \e A\dap M'N',
  \mbox{\ $\Ra$ by part 1)}\\
&\Lra& \exists M',N',B,C \. M\rast M' \e N\rast N' \e\\
&& A=BC \e
       B\dap M' \e C\dap N',\\
&&   \mbox{\ $\La$ by part 1), as $M'$ is passive}\\
&\Lra&  \exists B,C \. A=BC \e B\ap M \e C\ap N .
\EEQN

From the direction $\Ra$ follows:
$\Sae{MN} \sub \fl{\Sae{M}}\: (\Sae{N})$.\\
The direction $\La$ gives:
\EQN
\fl{\Sae{M}}\:(\Sae{N}) &=&
   \fl{\bigcup_{B\apind M} \Se{B}}\: (\bigcup_{C\apind N} \Se{C}) \\
&=& \bigcup_{B\apind M} \bigcup_{C\apind N} \Se{BC},
  \mbox{\ by continuity} \\
&\sub& \Sae{MN}, \mbox{\ from $\La$.}
\EEQN
\EDLE
\eproof

\LEMMA \label{passiveincl}
If $M\in \T_\s$ is passive and $a\in \Sae{M}$, then $a\ips M$.
\ELEMMA
\proof by structural induction on $a$.
\LP
\BU $\s=\t+\r$:
\\ The lemma is clear for $a=0$ and $a=1$.\\
Now let $a=(0,a')$.
As $M$ is passive, $M$ will not reduce to the form $\0 M'$ or $\1 M'$.
Therefore $\Outz(M) = \case\:M(\l y.y)\:\O$ is passive, too.
\EQN
a' &\in& \case\:(\Sae{M})\:(\Sae{\l y.y})\:(\Sae{\O}),
  \mbox{\ as\ } a \in \Sae{M} \\
&=& \Sae{\case\:M(\l y.y)\:\O}, \mbox{\ by Lemma \ref{constant}}\\
&=& \Sae{\Outz(M)}.
\EEQN
By the \ih we get $a' \ipt \Outz(M)$.\\
The case $a=(1,a')$ is analogous.
\LP
\BU $\s=\t\p\r$ is like $\s=\t+\r$.
\LP
\BU $\s=\t\f\r$:\\
Let $a=(X,a')$.\\
Let $N\in \T_\t$ and $X\ipt N$.
Then $MN$ is passive by Lemma \ref{passive}, 2).\\
$(X,a')\in \Sae{M}$ and $X\sub \Sae{N}$ imply
\[ a'\in \fl{\Sae{M}}\:(\Sae{N}) = \Sae{MN},
  \mbox{\ by Lemma \ref{passive}, 3)}.\]
By \ih we get $a'\ipr MN$.\\
Thus we have shown $a\ips M$.
\eproof

\medskip
We need a special lemma for \pcase\ giving its properties with respect
to the inclusive predicates.
It must be proved by induction on primes.
Note that such a lemma is not necessary for the other constants.

\LEMMA \label{pcase}
\mbox{ }
\DLE{3)}
\item[1)]
If $0\in \Sae{M_0}$ and $a\ips M_1$, then $a\ips \pcase\:M_0 M_1 M_2$.
\item[2)]
If $1\in \Sae{M_0}$ and $a\ips M_2$, then $a\ips \pcase\:M_0 M_1 M_2$.
\item[3)]
If $a\ips M_1$ and $a\ips M_2$, then $a\ips \pcase\:M_0 M_1 M_2$.
\EDLE
\ELEMMA
\proof We abbreviate $M=\pcase\:M_0 M_1 M_2$.
\DL{3)}
\item[1)]
The proof is by structural induction on $a$.\\
If $M_0\rast\0 M_0'$ for some $M_0'$, then $M\rast M_1$, and $a\ips M$
follows from Lemma \ref{converge}.\\
We assume in the following that not $M_0 \rast \0 M_0'$.
(Also $M_0\rast\1 M_0'$ is not possible because of $0\in \Sae{M_0}$.)\\
We give a case analysis on $a$:

\LP
\BU $\s=\t+\r:$\\
Let $a=(0,a'):$
\DLE{b)}
\item[a)]
We assume $M_1\rast\0 M_1'$ and $M_2\rast\0 M_2'$ for some $M_1',M_2'$.\\
Then $M\rast \0\:(\pcase\:M_0 M_1' M_2')$.\\
$(0,a')\ips M_1$ implies $a'\ipt \Outz(M_1)$.\\
From Lemma \ref{converge} and $\Outz(M_1)\rast M_1'$ follows
$a'\ipt M_1'$.\\
The \ih gives $a' \ipt \pcase\:M_0 M_1' M_2'$.\\
Therefore $a'\in \Sae{\pcase\:M_0 M_1' M_2'}$ and
\smallskip
\EQN
(0,a') &\in& 0\:(\Sae{\pcase\:M_0 M_1' M_2'})\\
&=& \Sae{\0\:(\pcase\:M_0 M_1' M_2')}, \mbox{\ by Lemma \ref{constant}}\\
&=& \Sae{M}, \mbox{\ by Lemma \ref{conv}}.
\EEQN

Furthermore $a'\ipt \Outz(M)$, as
$\Outz(M) \rast \pcase\:M_0 M_1' M_2'$, by Lemma \ref{converge}.

\item[b)]
We assume that {\em not} $(M_1\rast\0M_1' \et M_2\rast\0 M_2')$
for any $M_1',M_2'$.\\
Together with the assumption (not $M_0\rast \0 M_0'$) there is no reduct
of $M$ that is a redex. Then
\[ a\in \spcase\:(\Sae{M_0})\:(\Sae{M_1})\:(\Sae{M_2}) = \Sae{M},
  \mbox{\ by Lemma \ref{constant}}. \]
$M$ is passive (note that $M_1,M_2$ are not of functional type).
By Lemma \ref{passiveincl} we get $a \ips M$.
\EDLE

The case $a=0$ is contained in the proof for $a=(0,a')$,
and the cases $a=1$, $a=(1,a')$ are analogous.

\LP
\BU $\s=\t\p\r$ is like $\s=\t+\r$.

\LP
\BU $\s=\t\f\r$: Let $a=(X,a')$.\\
With the assumption (not $M_0\rast \0 M_0'$) there is no reduct of $M$
that is a redex. Then
\[ a\in \spcase\:(\Sae{M_0})\:(\Sae{M_1})\:(\Sae{M_2}) = \Sae{M},
\mbox{\ by Lemma \ref{constant}}. \]
It remains to show: $\forall N\in \T_\t\. (X\ipt N \Ra a'\ipr MN)$.\\
It is $MN = \pcase\:M_0 M_1 M_2 N \f \pcase \:M_0 (M_1 N) (M_2 N)$.
We get:
\EQN
X\ipt N &\Ra& a'\ipr M_1 N, \mbox{\ as \ } (X,a')\ips M_1\\
&\Ra& a' \ipr \pcase\: M_0 (M_1 N) (M_2 N), \mbox{\ by \ih}\\
&\Ra& a' \ipr MN, \mbox{\ by Lemma \ref{converge}.}
\EEQN
This concludes part 1) of the lemma.

\item[2)]
Part 2) is analogous to part 1).

\item[3)]
The proof is by structural induction on $a$.\\
If $M_0\rast \0 M_0'$ for some $M_0'$, then $M\rast M_1$, and $a\ips M$
follows from Lemma \ref{converge}.\\
If $M_0\rast \1 M_0'$ for some $M_0'$, then $M\rast M_2$, and again $a\ips M$.\\
We assume in the following that neither $M_0\rast \0 M_0'$
nor $M_0\rast \1 M_0'$.
We give a case analysis on $a$:

\LP
\BU $\s=\t+\r:$\\
Let $a=(0,a')$.
\DLE{b)}
\item[a)]
We assume $M_1\rast \0 M_1'$ and $M_2\rast\0 M_2'$ for some $M_1',M_2'$.\\
Then $M\rast \0\:(\pcase\:M_0 M_1' M_2')$.\\
From $a\ips M_1$, $a\ips M_2$ we conclude by Lemma \ref{converge}
that $a'\ipt M_1'$ and $a'\ipt M_2'$.\\
By \ih $a'\ipt \pcase\:M_0 M_1' M_2'$.
As in part 1) we conclude $a\ips M$.
\item[b)]
We assume that not ($M_1\rast \0 M_1'$ and $M_2\rast \0 M_2'$)
for any $M_1', M_2'$.
\\ As in part 1) we conclude $a\ips M$.
\EDLE
The case $a=0$ is contained in the proof for $a=(0,a')$,
and the cases $a=1$, $a=(1,a')$ are analogous.

\LP
\BU $\s=\t\p\r$ is like $\s=\t+\r$.

\LP
\BU $\s=\t\f\r:$\\
The argumentation is just the same as in part 1), except that we
conclude: \\
$X\ipt N \Ra a'\ipr M_1 N \et a'\ipr M_2 N$.
\eproof
\EDL

In the following lemma we collect all the properties of the relations
$\ips$ on elements of $D_\s$ that we need in the proof of the
Approximation Theorem.

\LEMMA[Inclusive Predicate Lemma] \label{inclusive}
In the following $d$ is an element of $D_\s$, $D_\t$, or $D_\r$, 
and $M,N \in \T_\s$.
\DLE{10)}
\item[1)]
$\bot \ips M$.
\item[2)] $\s=\t+\r:$
\ARR{lcl}
0 d \ip{\t+\r} M &\Lra& 0 d\sub \Sae{M} \et d\ipt \Outz(M)\\
1 d \ip{\t+\r} M &\Lra& 1 d\sub \Sae{M} \et d\ipr \Outi(M)
\EARR
\item[3)] $\s=\t\p\r:$
\EQN
d\ip{\t\p\r} M &\Lra& d\sub \Sae{M} \et \\
&& \sfst\: d \ipt \fst\:M \et \ssnd\:d \ipr \snd\:M
\EEQN
\item[4)] $\s=\t\f\r:$
\EQN
d \ip{\t\f\r} M &\Lra& d\sub \Sae{M} \et\\
&& \forall e\in D_\t, N\in \T_\t \.
     (e\ipt N \Ra \fl{d} e \ipr MN)
\EEQN
\item[5)]
Let $n\geq 0$ and $c$ be a constant of type $\s = \t_1\f\ldots \f \t_n \f \r$,
such that there is no reduction rule for $c$ with less than $n$
arguments.
Then $\Sb{c}\ips c$ iff
\[ d_i\ip{\t_i} M_i \mbox{\ for\ } 1\leq i\leq n
  \Ra (\Sb{c})d_1 \ldots d_n \ipr c M_1\ldots M_n .\]
\item[6)]
If $d\ips M$ and $M \downarrow N$, then also $d\ips N$.
\item[7)]
If $M\in \T_\s$ is passive and $d\sub\Sae{M}$, then $d\ips M$.
\item[8)]
If $0\in \Sae{M_0}$ and $d\ips M_1$, then $d\ips \pcase\:M_0 M_1 M_2$.
\item[9)]
If $1\in \Sae{M_0}$ and $d\ips M_2$, then $d\ips \pcase\:M_0 M_1 M_2$.
\item[10)]
If $d_1\ips M_1$ and $d_2\ips M_2$, then $d_1\cap d_2 \ips \pcase\:M_0 M_1M_2$.
\EDLE
\ELEMMA

Note: The parts 6) and 7) of this lemma replace the Lemma 5 of the
proof of the Approximation Theorem for the untyped $\l$-calculus in
\cite{Mosses/Plotkin}. A condition for the recursively typed $\l$-calculus
corresponding to that of Lemma 5 would be too complicated.

\proof
1), 2), and 3) are simple consequences of the definition of $\ips$.
\DLE{5)}
\item[4)]
\DLE{$\Ra:$}
\item[$\Ra:$]
$d\sub \Sae{M}$ is clear.\\
Now let $e\in D_\t$, $N\in \T_\t$ and $e\ipt N$.\\
Let $a\in \fl{d} e$. Then there is $X\sub e$ with $(X,a)\in d$.\\
From $(X,a)\ip{\t\f\r} M$ and $X\ipt N$ follows $a\ipr MN$.
\item[$\La:$]
Let $(X,a)\in d$. We show: $\forall N\. X\ipt N \Ra a\ipr MN$.\\
Let $e=\downc{X}$.
By Lemma \ref{leq} we get $e\ipt N$.
Then $a\in \fl{d} e \ipr MN$.
\EDLE
\item[5)]
The proof is by induction on $n$.
Note that $\r$ may be a functional type that varies with $n$.
$n=0$ is clear.\\
Now assume the proposition for $c$ is true for some $n\geq 0$;
we prove it for $n+1$:
\ARR{lcl}
\Sb{c}\ips c \\
 \mbox{iff\ }
d_i\ip{\t_i}M_i \mbox{\ for\ } 1\leq i\leq n &\Ra&
(\Sb{c}) d_1\ldots d_n \ip{\t_{n+1}\f\r} c M_1\ldots M_n,\\
&& { \mbox{by \ih}}\\
\mbox{iff\ } d_i\ip{\t_i}M_i \mbox{\ for\ } 1\leq i\leq n &\Ra&
(\Sb{c}) d_1\ldots d_n \sub \Sae{c M_1\ldots M_n} \et    \\
&& (d_{n+1}\ip{\t_{n+1}} M_{n+1} \Ra \\
&&   (\Sb{c}) d_1\ldots d_{n+1} \ipr c M_1\ldots M_{n+1}),\\
&& {\mbox{by part 4).}}
\EARR
Lemma \ref{constant} says
$\Sae{c M_1\ldots M_n} = (\Sb{c})\:(\Sae{M_1})\ldots(\Sae{M_n})$,
therefore $(\Sb{c}) d_1\ldots d_n \sub \Sae{c M_1\ldots M_n}$
is fulfilled.
\item[6)]
Follows from Lemma \ref{converge}.
\item[7)]
Follows from Lemma \ref{passiveincl}.
\item[8), 9) and 10)] follow from Lemma \ref{pcase}.
\eproof
\EDLE

\medskip
The Approximation Theorem would be proved if we could show that
$\Se{M} \ips M$ for all $M\in \T_\s$.
We will now prove, by structural induction on $M$, a stronger
statement in order to handle free variables in the case of abstraction.

\LEMMA[Approximation Lemma] \label{approx}
Let $M\in \T_\s$, $\ep\in \Env$,
$x_i^{\s_i}$ $(1\leq i\leq n, n\geq 0)$ be a sequence of
distinct variables, $d_i\in D_{\s_i}$ and $N_i\in \T_{\s_i}$ for all $i$.\\
If $d_i \ip{\s_i} N_i$ for all $i$, then
\[ \S{M}(\ep[x_1\mto d_1,\ldots,x_n\mto d_n]) \ips
   M[x_1\subst N_1,\ldots,x_n\subst N_n]. \]
Here $\ep[x_1\mto d_1,\ldots,x_n\mto d_n]$ is the environment that maps
$x$ to $\ep(x)$ if $x\neq x_i$ for all $i$, and $x_i$ to $d_i$.
$M[x_1\subst N_1,\ldots,x_n\subst N_n]$ is the result of the simultaneous
substitution of the $N_i$ for the free occurrences of $x_i$ in $M$,
with appropriate renaming of bound variables of $M$.
\ELEMMA

\proof by structural induction on $M$.\\
For any $\ep'\in\Env$ we abbreviate 
$\tild{\ep'} = \ep'[x_1\mto d_1,\ldots,x_n\mto d_n]$,
and for any term $L$ we write $\tild{L} = L[x_1\subst N_1,\ldots,x_n\subst
 N_n]$.\\
We cite the parts of the Inclusive Predicate Lemma simply by part i).
The use of parts 1) -- 5) should be obvious and is often not mentioned.
\LP
\BU $M=\O$: $\S{\O}\tile =\bot \ips \O$.

\LP
\BU $M=\0,\; \s=\t\f(\t+\r):$\\
To show $\S{\0}\tile \ips \0$, we prove
$d\ipt N \Ra 0 d \ip{\t+\r} \0 N$.\\
We have $0d\sub 0(\Sae{N}) = \Sae{\0 N}$.
Furthermore $d\ipt \Outz(\0 N)$ by part 6),
as $\Outz(\0 N)\rast N$.

\LP
\BU $M=\1$ is analogous.

\LP
\BU $M=\case,\; \s = (\t+\r)\f(\t\f\psi)\f(\r\f\psi)\f\psi :$\\
To show $\S{\case}\tile \ips \case$, we have to prove:
\[ d_0\ip{\t+\r}M_0 \e d_1\ip{\t\f\psi}M_1 \e d_2\ip{\r\f\psi}M_2 \Ra
  \scase\:d_0 d_1 d_2 \ipp \case\:M_0 M_1 M_2 .\]
This is clear for $d_0=\bot$.\\
Now let $d_0= 0 d_0'$.
\DLE{b)}
\item[a)] We assume $M_0 \rast \0 M_0'$ for some $M_0'$.\\
As $d_1\ip{\t\f\psi}M_1$ and $d_0' \ipt \Outz(M_0)$, we get
\[ \scase\:d_0 d_1 d_2 = \fl{d_1}d_0' \ipp M_1(\Outz(M_0)).\]
We have $\case\:M_0 M_1 M_2 \rast M_1 M_0'$
and $M_1(\Outz(M_0))\rast M_1 M_0'$,\\
so $\scase\:d_0 d_1 d_2 \ipp \case\:M_0 M_1 M_2$ by part 6).

\item[b)] We assume that not $M_0\rast \0 M_0'$ for any $M_0'$.\\
$M_0\rast \1 M_0'$ is also impossible.
So there is no reduct of $\case\:M_0 M_1 M_2$ that is a redex.\\
From Lemma \ref{constant} we conclude:
\EQN
\scase\:d_0 d_1 d_2 &\sub& \scase\:(\Sae{M_0})\:(\Sae{M_1})\:(\Sae{M_2}) \\
&=& \Sae{\case\:M_0 M_1 M_2} .
\EEQN

Furthermore $\case\:M_0 M_1 M_2$ is passive,
and $\scase\: d_0 d_1 d_2 \ipp \case\:M_0 M_1 M_2$ follows from part 7).
\EDLE
The case $d_0=1d_0'$ is analogous.

\LP
\BU $M=\pcase, \; \s=(\t+\r)\f\psi\f\psi\f\psi :$\\
We have to prove:\\
$ d_0\ip{\t+\r} M_0 \e d_1\ipp M_1 \e d_2\ipp M_2 \Ra
   \spcase\:d_0 d_1 d_2 \ipp \pcase\:M_0 M_1 M_2 .$\\
For $d_0=\bot$ we have $\spcase\:d_0 d_1 d_2 = d_1 \cap d_2$.
The result follows from part 10).\\
For $d_0=0d_0'$ we use part 8), for $d_0=1d_0'$ part 9).

\LP
\BU $M=\pair$ is like $M=\0$.

\LP
\BU $M=\fst, \; \s=(\t\p\r)\f\t :$\\
$d\ip{\t\p\r}N \Ra \sfst\: d \ipt \fst\:N$
follows directly from part 3).

\LP
\BU $M=\snd$ is analogous.

\LP
\BU $M=x :$\\
If $x=x_i$ for some $i$, then
$\S{x}\tile = d_i \ips N_i = \tild{x}$.\\
Now let $x\neq x_i$ for all $i$.
Then $\S{x}\tile = \ep(x) \sub \Sae{x}$.
$x$ is passive. From part 7) follows $\S{x}\tile \ips x$.

\LP
\BU $M=NP$, where $N:\t\f\s$ and $P:\t$:\\
By \ih we have $\S{N}\tile \ip{\t\f\s} \tild{N}$
and $\S{P}\tile \ipt \tild{P}$.\\
Therefore $\fl{\S{N}\tile}\:(\S{P}\tile) \ips \tild{N}\;\tild{P}$,
by part 4).\\
Thus we get $\S{NP}\tile \ips \tild{NP}$.

\LP
\BU $M=\l x^\t.M', \; \s=\t\f\r :$\\
We may assume that $x$ is no $x_i$ and $x$ does not occur free in
any $N_i$. ($x$ can be renamed by $\alpha$-conversion.)\\
First we prove that $\S{\l x.M'}\tile \sub \Sae{\tild{\l x.M'}}$.
\EQN
\S{\l x.M'}\tile &=& \Pr(d\in D_\t \mto \S{M'}(\tile[x\mto d])) \\
&=& \Pr(d\in D_\t \mto \S{M'}(\tild{\ep[x\mto d]})), \mbox{\ as
  $x$ is no $x_i$} \\
&\sub& \Pr(d \in D_\t \mto \Sa{\tild{M'}}(\ep[x\mto d])),\\
&& \mbox{\ as\ } \S{M'}(\tild{\ep[x\mto d]}) \ipe{\ep[x\mto d]}{\r}
   \tild{M'} \mbox{\ by \ih} \\
&=& \Pr(d\in D_\t \mto \bigcup_{A\apind \tild{M'}} \S{A}(\ep[x\mto d]))\\
&=& \bigcup_{A\apind \tild{M'}}
     \Pr(d\in D_\t \mto \S{A}(\ep[x\mto d])) \\
&=& \bigcup_{A\apind \tild{M'}} \Se{\l x.A}\\
&=& \bigcup_{B\apind \tild{\l x.M'}} \Se{B},\\
&& \mbox{\ as\ } A\ap \tild{M'} \Lra
     \l x.A \ap \l x.\tild{M'} = \tild{\l x.M'},
     \mbox{\ since $x$ is no $x_i$} \\
&=& \Sae{\tild{\l x.M'}}
\EEQN
Now we prove that:
$d\ipt N \Ra \fl{\S{M}\tile}\: d \ipr \tild{M}N$.
\EQN
\fl{\S{M}\tile}\: d &=& \S{M'}(\tile[x\mto d]) \\
&=& \S{M'}(\ep[x_1\mto d_1,\ldots,x_n\mto d_n, x\mto d]),
    \mbox{\ as $x$ is no $x_i$} \\
&\ipr& M'[x_1\subst N_1,\ldots,x_n\subst N_n, x\subst N],
    \mbox{\ by \ih}
\EEQN
Furthermore we have:
\EQN
\tild{M}N &=& (\l x.\tild{M'}) N, \mbox{\ as $x$ is no $x_i$}\\
&\f& (M'[x_1\subst N_1,\ldots,x_n\subst N_n])[x\subst N]\\
&=& M'[x_1\subst N_1,\ldots,x_n\subst N_n, x\subst N],
   \mbox{\ as $x$ is not free in any $N_i$}
\EEQN
From part 6) follows $\fl{\S{M}\tile}\: d \ipr \tild{M}N$.
\eproof

\medskip\LP
{\bf Proof of the Approximation Theorem:}\\
$\Se{M} \sub \Sae{M}$ follows from $\Se{M} \ips M$,
which holds by the preceding lemma.
\eproof

\begin{corollary}
For all terms $M$ and environments $\ep$:
\[ \Se{M} = \bigcup \set{\Se{A} \st A \mbox{\ is a normal form}
  \et \exists N\. M\rast N \e A\pre N} \]
\end{corollary}
\proof
$\Sae{M} \sub$ the right-hand side,
and the right-hand side $\sub \Se{M}$.
\eproof
\LP
Note: The original paper \cite{Wadsworth} gives a definition
of approximations in the form of this corollary,
for the untyped $\l$-calculus.

\begin{corollary}
The semantics of the fixed point combinator\\
$Y_\s = \l y^{\s\f\s}. (\l x.y(x x))(\l x.y(x x))$ is
\[\Se{Y_\s} = \Pr(f\in D_{\s\f\s} \mto \bigcup_{n\geq 0} f^n(\bot)),\]  
so $\fl{\Se{Y_\s}} f$ is the least fixed point of $\fl{f}$.
\end{corollary}
\proof
The approximations of $Y_\s$ are just the terms $\l y.y^n \O$,
with $y^0\O=\O$ and $y^{n+1}\O=y(y^n\O)$.
\EQN
\Se{Y_\s} &=& \Sae{Y_\s}\\
&=& \bigcup_{n\geq 0} \Se{\l y.y^n\O} \\
&=& \bigcup_{n\geq 0} \Pr(f\in D_{\s\f\s} \mto f^n(\bot)) \\
&=& \Pr(f\in D_{\s\f\s}\mto \bigcup_{n\geq 0} f^n(\bot))
\EEQN
\eproof

\medskip
Let us continue our discussion of the definition of approximations.
In the case of the untyped $\l$-calculus \cite{Barendregt} it is
possible to define least approximations via head normal forms.
Let us look at this approach more abstractly:
We are given a set $H$ of normal forms with the property:
If $A\in H$ and $A\pre M$, then $A\dap M$.
This means that an $H$-prefix of a term $M$ does not change by
reductions of $M$.
In the case of the untyped $\l$-calculus $H$ is the set consisting
just of $\O$ and all terms of the form 
$\l x_1\ldots x_n. y A_1\ldots A_m$ with $A_i\in H$.
We define
\[ \SH{M}\ep = \bigcup \set{\Se{A} \st A\in H \et \exists N\.
   M\rast N \e A\pre N}. \]
$H$ should fulfill: $\SH{M}\ep = \Se{M}$ for all $M,\ep$.
We show that a set $H$ with this property and the property above
does not exist for our calculus with
\pcase:\\
Let $M=\pcase\:x\,\0\, \O$.
It is $M\not\in H$, because of the first property of
$H$ and as not $M\dap \pcase\:x\,\0\,\0$.
For all $A\pre M$ with $A\neq M$ we have $\S{A}(\bot[x\mto 0]) =\bot$.
Therefore $\SH{M}(\bot[x\mto 0]) =\bot \neq 0 = \S{M}(\bot[x\mto 0])$.

\LP
Let us now consider the sequential calculus {\em without} \pcase.
In this case we can define two sets $H$ with the desired properties.

\DEFI
A normal form $A$ is a {\de minimum normal form} (mnf) iff
for all $B\pre A$: $\S{B}=\S{A} \Ra B=A$.\\
A normal form $A$ is a {\de constant normal form} (cnf) iff
\[A=\O \mbox{\ or\ } A=\l x_1\ldots x_n. y A_1\ldots A_m,\] where $n\geq 0$,
$m\geq 0$, $y$ is a variable or a constant $\not\in \set{\O,\pcase}$,
the $A_i$ are cnfs and for $y\in \set{\fst,\snd,\case}$
and $m\geq 1$ it is $A_1\neq \O$.
\EDEFI
Constant normal forms resemble the normal forms of $H$ defined
by head normal forms above, for the untyped $\l$-calculus.

\LEMMA \label{mnfcnf}
Every minimum normal form without \pcase\ is a constant normal form.
\ELEMMA
\proof
Suppose $A$ is a normal form without \pcase\ that is no cnf.
We show by structural induction on $A$ that $A$ is no mnf.
\\ We have $A= \l x_1\ldots x_n. y A_1\ldots A_m$, $n\geq 0$,
$m\geq 0$, $y$ a variable or a constant, and one of the following
three cases:
\DLE{3)}
\item[1)] $y=\O$ and ($n>0$ or $m>0$).\\
Then $\O\pre A$, $\O\neq A$ and $\S{\O}=\S{A}$, so $A$ is no mnf.
\item[2)] Some $A_i$ is no cnf.\\
By \ih $A_i$ is no mnf. Then also $A$ is no mnf.
\item[3)] $y$ is \fst, \snd\ or \case\ and $A_1=\O$.\\
Then $\S{A}=\S{\O}$, $A$ is no mnf.
\eproof
\EDLE

\LEMMA \label{cnfapp}
If $A$ is a constant normal form and $A\pre M$, then $A\dap M$.
\ELEMMA
\proof
We prove: If $A$ is a cnf, $A\pre M$ and $M\f N$, then $A\pre N$,
by structural induction on $A$. (The lemma follows by simple
induction on reductions $M\rast N$.)\\
The case $A=\O$ is clear.\\
Now let $A=\l x_1\ldots x_n. y A_1\ldots A_m$.
Then $M=\l x_1\ldots x_n. y M_1\ldots M_m$ with $A_i\pre M_i$ for all $i$.\\
The term $y M_1\ldots M_m$ is no redex:\\
This is clear if $y$ is a variable or \0, \1, or \pair.\\
If $y=\fst$ or $y=\snd$, and $m\geq 1$, then $A_1\neq \O$ 
and $A_1$ is not of the form $\pair\:A' A''$.
So $M_1$ is not of this form either.\\
If $y=\case$ and $m\geq 1$, then $A_1\neq \O$
and $A_1$ and $M_1$ are not of the form $\0 A'$ or $\1 A'$.\\
Thus there is some $j$ with $M_j\f N_j$ and
$N= \l x_1\ldots x_n. y M_1\ldots M_{j-1} N_j M_{j+1}\ldots M_m$.
By the \ih we get $A_j\pre N_j$, therefore $A\pre N$.
\eproof

\medskip
By this lemma the set of cnfs (and the set of mnfs) has the first
of the two properties of $H$.
We define two new approximation sets for terms:
\EQN
\B{M} &=& \set{A \st A \mbox{\ is a mnf}\et \exists N\. M\rast N \e A\pre N}\\
\C{M} &=& \set{A \st A \mbox{\ is a cnf}\et \exists N\. M\rast N \e A\pre N}
\EEQN
For the sequential calculus without \pcase\ we have:
\[ \B{M} \sub \C{M} \sub \A{M}. \]
The first inclusion   follows from Lemma \ref{mnfcnf},
the second from Lemma \ref{cnfapp}.\\
$\B{M}\sub \A{M}$ is not valid for $M=\pcase\:x\,\0\,\0$:
We have $\pcase\:x\,\0\,\O \in \B{M}$,
but $\pcase\:x\,\0\,\O \not\in \A{M}$.\\
In every case, also for \pcase:
\[ \Sae{M} = \bigcup_{A\in \A{M}} \Se{A} \sub 
   \bigcup_{A\in \B{M}} \Se{A} \mbox{\ for all\ } \ep\in \Env. \]
This is because for every normal form $A$ there is a mnf $B\pre A$
with $\S{A}=\S{B}$.

\LP
We combine these results with the Approximation Theorem:
\THEOREM
In the sequential calculus without \pcase: For all terms $M$
and environments $\ep$,
\[ \bigcup_{A\in \B{M}} \Se{A} = \bigcup_{A\in \C{M}} \Se{A} =
   \Sae{M} = \Se{M} .\]
\ETHEOREM
With this theorem the set of mnfs and the set of cnfs both have
the second property of $H$.

[My conjecture is that in the sequential calculus $\B{M}$ is the
least approximation of $M$ with the same semantics as $M$.
More precisely the conjecture is: Let $I$ be an ideal of normal forms
such that for all $A\in I$ there is $N$ with $M\rast N$ and $A\pre N$,
and $\S{M}=\bigcup_{A\in I} \S{A}$. Then $\B{M}\sub I$.]

\section{Adequacy and full abstraction}

The classical semantical analysis of the programming language PCF
\cite{Plotkin}
proceeds as follows: The closed terms of the ground type integer are
singled out as {\em programs}. Programs are regarded as the only terms
whose syntactical values (integers) can be observed directly.
All other terms must be observed through program contexts.
If the semantics of a programm $M$ is an integer value $i$,
then $M$ can be reduced to $i$. This result is called the {\em adequacy} of
the semantics.
Then an operational preorder is defined on terms: $M\suq N$ iff for
all contexts $C[\:]$ such that $C[M]$ and $C[N]$ are programs,
if $C[M]\rast i$, then also $C[N]\rast i$.
If $\S{M}\sub \S{N}$, then $M\suq N$; this follows from soundness and
adequacy. The converse, {\em full abstraction}, is not true for sequential
PCF, but holds for PCF with a parallel conditional.

We follow the same programme for our recursively typed $\l$-calculus.
We choose the closed terms of type $\bool = \void +\void$ as our programs.
Thus the observable non-bottom syntactical values are the terms of the form
$\0M$ or $\1 M$. We have chosen the smallest type with more than
one element. (Any non-functional, non-trivial type, built from
$+$ and $\p$ only, would do as well.)

\DEFI
The set of {\de programs} is $\Prog = \Tc{\bool}$.\\
We define the {\de operational evaluation function}
$\Oper : \Prog \f D_{\scriptsize\bool}$ by 
$\Op{M} = 0 $ if $M\rast \0M'$,
$\Op{M}= 1$ if $M\rast \1M'$, for some $M'$,
and $\Op{M} = \bot$ otherwise.
\EDEFI
We want to prove adequacy (that the reduction of a program reaches its
semantic value) from the Approximation Theorem of the preceding chapter.
We need the following lemma:
\LEMMA
Let $\s\in\Tinf$ and $A\in \N_\s$ be a normal form with $\Sb{A}\neq \bot$.\\
If $\s=\t+\r$, then $A=\0 A'$ or $A=\1A'$ for some $A'$.\\
If $\s=\t\p\r$, then $A=\pair\:A'A''$ for some $A',A''$.
\ELEMMA
\proof by structural induction on $A$.\\
We suppose $A$ is of type $\t+\r$ or $\t\p\r$.
Then $A= c A_1\ldots A_n$, $n\geq 0$, with $c$ a constant and the
$A_i$ normal forms. We give a case analysis on $c$:
\LP
$c=\0, \1$ or \pair: The lemma is fulfilled.
\LP
$c=\fst$ or \snd:\\
Then $n\geq 1$. 
$\Sb{A}\neq \bot$ implies $\Sb{A_1}\neq \bot$
implies $A_1=\pair\: A'A''$ by induction hypothesis. Then $A$ is no normal form.
\LP
$c=\case:$\\
Then $n\geq 3$.
$\Sb{A}\neq \bot$ implies $\Sb{A_1}\neq \bot$
implies $A_1=\0A_1'$ or $A_1=\1A_1'$ by induction
hypothesis. Then $A$ is no normal form.
\LP
$c=\pcase:$\\
Then $n=3$.
If $\Sb{A_1}\neq \bot$, then $A_1=\0A_1'$ or $A_1=\1A_1'$ by \ih and
$A$ is no normal form.\\
If $\Sb{A_1}=\bot$, then $\Sb{A}=\Sb{A_2}\cap \Sb{A_3} \neq \bot$.\\
If $\s=\t+\r$, then by \ih either ($A_2=\0A_2'$, $A_3=\0A_3'$)
or ($A_2=\1A_2'$, $A_3=\1A_3'$). In both cases $A$ is no normal form.\\
If $\s=\t\p\r$, then by \ih $A_2=\pair\:A_2'A_2''$ and
$A_3=\pair\:A_3'A_3''$ and $A$ is no normal form.
\eproof
\THEOREM[Adequacy]
For all $M\in \Prog$: $\Op{M}=\Sb{M}$.
\ETHEOREM
\proof
$\Op{M}\sub \Sb{M}$ follows from soundness:
If $M\rast \0M'$, then $\Sb{M}=\Sb{\0M'}=0$;
and if $M\rast \1M'$, then $\Sb{M}=\Sb{\1M'}=1$.\\
It remains to show the adequacy: $\Sb{M}\sub \Op{M}$.\\
Suppose $\Sb{M}=0$. By the Approximation Theorem there is an
approximation $A\ap M$ with $\Sb{A}=0$. From the preceding lemma follows
$A=\0A'$ for some $A'$, therefore $\Op{M}=0$.
Analogously $\Sb{M}=1$ implies $\Op{M}=1$.
\eproof

Note that this theorem is also valid for the sequential calculus without
\pcase. It can also be proved directly using the inclusive predicate
technique, with a proof a bit easier than the proof of the
Approximation Theorem, \eg the passive terms are not needed.

Now we define the operational preorder on terms, based on the observation of
terms through program contexts.
\DEFI Let $M,N \in \T_\s$.
$M\suq N$ iff for all contexts $C[\:]$, such that $C[M]$ and $C[N]$ are
programs, $\Op{C[M]} \sub \Op{C[N]}$ holds.
\EDEFI
\THEOREM[Full abstraction]
For all $M,N\in \T_\s$: $M\suq N$ iff $\S{M}\sub\S{N}$.
\ETHEOREM
The direction ``If $\S{M}\sub\S{N}$ then $M\suq N$'' follows easily from
soundness and adequacy:
$\Op{C[M]} = \Sb{C[M]} \sub \Sb{C[N]} = \Op{C[N]}$.
This holds also for the sequential calculus without \pcase.
In this case the contexts are restricted. Therefore the opposite direction
is not valid for the sequential calculus, as can be shown by the same
example as in \cite{Plotkin}.

For the proof of the opposite direction (for the parallel calculus)
we prove a lemma that states the definability of all finite elements
of the semantics.
\LEMMA[Definability]
For all finite $d\in D_\s$ there is a closed term $M\in \Tc{\s}$
with $\Sb{M}=d$.
\ELEMMA
We recall that finite elements are the elements that are downward closures
of finite sets of primes.
In our term construction we use the following parallel function \andf\
instead of \pcase:
\EQN
\andf &:& \bool\f\bool\f\bool,\mbox{\ defined as}\\
\andf &=& \l xy.\pcase\:x\,y\,\1.
\EEQN
Here and in the following we interpret the Boolean value $0$ as true
and $1$ as false, and chose the names of our functions accordingly.
(We made this choice in order to interpret \case\ like if-then-else,
with the second argument as true-part and the third argument as false-part.)
The semantics of \andf\ fulfills:
$(\Sb{\andf})00=0$,
$(\Sb{\andf})1\bot=1$,
$(\Sb{\andf})\bot 1=1$.
Here we show that all finite elements are definable from
\andf\ and the sequential constants.
In the next chapter we will show that also \pcase\ (which is not
finite) is definable from \andf.
\LP
\proof
We have to introduce some notions first.
A term $C: \bool$ is called a {\de condition} iff for every environment $\ep$:
\[ (\forall \ep' \sup \ep\. \S{C}\ep' \neq 0) \Ra \Se{C} =1 .\]
The semantics of a condition is so ``dense'' that it gives the value $1$
for every environment that cannot be enlarged to give the value $0$.\\
A {\de conditioned prime} is a pair $C\co a$ of a condition $C$ and a prime $a$.
In the course of our construction the condition of $C\co a$ will be used to
accumulate a term that checks function arguments.
The intuitive semantics of the ``mixed term'' $C\co a$ is the prime $a$
for every environment $\ep$ with $\Se{C}=0$.
For a set $P$ of primes, $\Cond(P)$ is the set of all conditioned
primes $C\co a$ with $a\in P$.\\
A set $X$ of conditioned primes is called {\de consistent}
iff for all $C\co a, C'\co a' \in X$ holds:
$(\exists \ep\. \Se{C}=\Se{C'}=0) \Ra a \con a'$.
\LP
For $M\in \T_\s$, $X\sub \Cond(P(\s))$ finite and consistent,
we define a predicate $\term$:
\[ M\term X \mbox{\ iff\ }
   \Se{M}=\downc{\set{a\st \exists C\. (C\co a)\in X \e \Se{C}=0}}
   \mbox{\ for all $\ep$.} \]
For $M\in \Tc{\s\f\bool}$, $X\sub P(\s)$ finite and consistent,
we define a predicate $\eq$:
\[ M\eq X \mbox{\ iff\ } \fl{\Sb{M}}\:d = \left\{
   \begin{array}{ll}
   0, & \mbox{if $X\sub d$}\\
   1, & \mbox{if $d\ncon X$}\\
   \bot & \mbox{otherwise}
   \end{array}
   \right. \]
where $d\ncon X$ means: $\exists a\in d, b\in X\. \mbox{not\ } a\con b$.

\medskip
\LP
We prove for every $n\geq 0$ and every $\s\in \Tinf$:
\DLE{\bf 2)}
\item[{\bf 1)}]
For every finite and consistent $X\sub \Cond(P_n(\s))$
there is $M\in \T_\s$ with $M\term X$.
\item[{\bf 2)}]
For every finite and consistent $X\sub P_n(\s)$
there is $M\in \Tc{\s\f\bool}$ with $M\eq X$.
\EDLE
\medskip
We use abbreviations for the following function terms:
\ARR{lcll}
\ifz &=& \l xyz.\case\:x\,(\l w.y)\,(\l w.z) : &\bool\f\s\f\s\f\s \\
\notf &=& \l x.\ifz\,x\,\1\,\0 : &\bool\f\bool \\
\orf &=& \l xy.\notf\,(\andf\,(\notf\,x)\,(\notf\, y))
    : & \bool\f\bool\f\bool
\EARR
The semantics of \orf\ is:
$(\Sb{\orf}) 1 1=1$,
$(\Sb{\orf}) 0 \bot=0$,
$(\Sb{\orf}) \bot 0=0$.
\medskip
\LP
The proof of statements 1) and 2) is by simultaneous induction on $n$:
\DLE{$n=0$:}
\item[$n=0$:]
1) $X=\emptyset$. $\O\term X$.\\
2) $X=\emptyset$. $(\l x.\0) \eq X$.
\EDLE
Induction step:
\LP
{\bf 1)}
Let $X\sub \Cond(P_{n+1}(\s))$ be finite and consistent.
We construct $M \term X$ by case analysis over $\s$.
\LP
\BU $\s=\void$: $X=\emptyset$, $\O\term X$.

\LP
\BU $\s=\t+\r$:\\
Define the condition sets 
$C^0 = \set{C\st \exists a\gep 0\. (C\co a)\in X}$ and\\
$C^1 = \set{C\st \exists a\gep 1\. (C\co a)\in X}$.\\
Define the term $M_0: \bool$ as $M_0=\1$ for $C^0=\emptyset$,
otherwise as $M_0 = \orf\:C^0_1\:(\orf\:C^0_2 \ldots C^0_j)$
for some enumeration $\set{C^0_1, C^0_2,\ldots, C^0_j} = C^0$.
Analogously, $M_1$ is defined as an \orf-term of the elements of $C^1$.\\
Let $X^0=\set{C\co a\st (C\co(0,a))\in X}$ and
    $X^1=\set{C\co a\st (C\co(1,a))\in X}$.  
It is $X^0\sub \Cond(P_n(\t))$ and $X^1\sub\Cond(P_n(\r))$,
both are finite and consistent.
By the \ih there are terms $N_0\in \T_\t$, $N_1\in \T_\r$
with $N_0\term X^0$ and $N_1\term X^1$.\\
We build the term 
\[ M= \ifz \:M_0\,(\0 N_0)\,(\ifz\:M_1\,(\1 N_1)\,\O) \]
and show that $M\term X$,
\\ \ie for all $\ep$,
$\Se{M} = \downc{Y}$ with
$Y= \set{a\st \exists C\. (C\co a)\in X \e \Se{C}=0}$:

\medskip
\LP  $\star \;\Se{M}\sub \downc{Y}:$\\
Let $a\in \Se{M}$. We show $a\in \downc{Y}$ in each of the
two cases:
\DLE{a)}
\item[a)]
$\Se{M_0}=0$: Then $a\in 0\,(\Se{N_0})$.\\
First let $a=0$. There is some $C\in C^0$ with $\Se{C}=0$.
$(C\co a')\in X$ for some $a'\gep 0$, therefore $0\in \downc{Y}$.\\
Now let $a=(0,a')$. Then $a'\in \Se{N_0}$.
Since $N_0 \term X^0$, there is $(C\co a'')\in X^0$
with $\Se{C}=0$ and $a'\lep a''$.
$(C\co (0,a''))\in X$, therefore $(0,a')\in \downc{Y}$.
\item[b)]
$\Se{M_0}=1$ and $\Se{M_1}=0$: Then $a\in 1\,(\Se{N_1})$.\\
Analogously to case a) we show that $a\in \downc{Y}$.
\EDLE
  
\medskip
\LP  $\star\;\Se{M}\sup \downc{Y}:$\\
Let $a\in Y$, \ie $(C\co a)\in X$ and $\Se{C}=0$ for some $C$.
We show $a\in \Se{M}$ in each of the four cases:
\DLE{b)}
\item[a)] $a=0$:\\
$C\in C^0$, therefore $\Se{M_0}=0$ and $0\in \Se{M}$.
\item[b)] $a=(0,a')$:\\
Again $C\in C^0$, therefore $\Se{M_0}=0$ and $\Se{M} = 0\,(\Se{N_0})$.
$(C\co a')\in X^0$, therefore $a'\in \Se{N_0}$, as $N_0 \term X^0$.
It follows $(0,a')\in \Se{M}$.
\item[c)] $a=1$:\\
Then $C\in C^1$, therefore $\Se{M_1}=0$.\\
We show that $\Se{M_0}=1$, \ie for all $C'\in C^0$: $\Se{C'}=1$.
Here we use the fact that $C'$ is a condition:\\
Let $\ep'\sup\ep$. Then $\S{C}\ep' = 0$.
$\S{C'}\ep' =0$ would contradict the consistency of $X$,
as $C\in C^1$ and $C'\in C^0$.
Therefore $\S{C'}\ep' \neq 0$.
We conclude $\Se{C'}=1$.\\
So we have $\Se{M_0}=1$, $\Se{M_1}=0$ and $1\in \Se{M}$.
\item[d)] $a=(1,a')$:\\
As in case c) we have $\Se{M_0}=1$, $\Se{M_1}=0$
and $\Se{M}= 1\,(\Se{N_1})$.\\
$(C\co a')\in X^1$, therefore $a'\in \Se{N_1}$, as $N_1 \term X^1$.
It follows $(1,a')\in \Se{M}$.
\EDLE

\bigskip
\LP \BU $\s=\t\p\r$:\\
Let $X^0=\set{C\co a\st (C\co (0,a))\in X} \sub \Cond(P_n(\t))$,
and 
    $X^1=\set{C\co a\st (C\co (1,a))\in X} \sub \Cond(P_n(\r))$.
Both sets are finite and compatible.\\
By the \ih there are terms $N_0,N_1$ with
$N_0\term X^0$ and $N_1\term X^1$. Let $M=(N_0,N_1)$.
\EQN
\Se{M} &=& \spair\:(\Se{N_0})\,(\Se{N_1}) \\
&=& \set{0}\cp \downc{\set{a\st \exists C\. (C\co a)\in X^0 \e \Se{C}=0}}
\cup \\
&& \set{1}\cp \downc{\set{a\st \exists C\. (C\co a)\in X^1 \e \Se{C}=0}}\\
&=& \downc{\set{a\st \exists C\. (C\co a)\in X \e \Se{C}=0}}
\EEQN

\medskip
\LP \BU $\s=\t\f\r$:\\
Let $X=\set{C_i\co (Y_i,a_i) \st 1\leq i \leq k}$ be an enumeration of the
elements of $X$.\\
For all $i$, $Y_i\sub P_n(\t)$ is finite and consistent.
By the \ih there is $N_i \eq Y_i$ for all $i$.\\
Let $x$ be a variable of type $\t$ that does not occur free in any $C_i$.
Let $D_i= \andf\:C_i\,(N_i x)$.
We define $Z=\set{D_i\co a_i\st 1\leq i\leq k}$ and first prove that
$Z\sub \Cond(P_n(\r))$ and $Z$ is consistent:

\medskip
\LP  $\star\; D_i=\andf\:C_i\,(N_i x)$ is a condition:\\
Let $\ep$ be an environment such that for all $\ep'\sup \ep$,
$\S{D_i}\ep' \neq 0$.
We have to show that $\Se{D_i}=1$.\\
Assume $\Se{C_i}\neq 1$.
As $C_i$ is a condition, there is $\ep''\sup \ep$ with $\S{C_i}\ep''=0$.
Let $\ep'=\ep''[x\mto \downc{Y_i}]$.\\
Then $\S{C_i}\ep'=0$, as $x$ does not occur free in $C_i$.
Furthermore $\S{N_i x}\ep'=0$, as $N_i\eq Y_i$.
Together we get $\S{D_i}\ep'=0$.\\
Then $\ep$ and $\ep'$ cannot have an upper bound.
(For such an upper bound $\delta$ would be: $\delta\sup \ep$
and $\S{D_i}\delta =0$.)
As $\ep''\sup \ep$, it must be $\ep(x)\ncon \ep'(x)=\downc{Y_i}$.
Hence $\Se{N_i x}=1$, and we conclude $\Se{D_i}=1$.

\medskip
\LP  $\star\; Z$ is consistent:\\
Let  $\Se{D_i}=\Se{D_j}=0$ for some $i,j,\ep$.
Then $\Se{C_i}=\Se{C_j}=0$, hence $(Y_i,a_i)\con (Y_j,a_j)$.
Also $\Se{N_i x}=\Se{N_j x}=0$, therefore $Y_i\sub \ep(x)$
and $Y_j\sub \ep(x)$.
So $Y_i\con Y_j$ and we conclude $a_i\con a_j$.

\medskip
We have proved that $Z\sub \Cond(P_n(\r))$ is a finite, consistent,
conditioned prime set.
By \ih there is $N\term Z$.
Let $M=\l x.N$.
We prove $M\term X$, \ie
\[ \Se{M}=\Pr(d\in D_\t \mto \S{N}(\ep[x\mto d])) =
   \downc{\set{(Y_i,a_i)\st 1\leq i\leq k \e \Se{C_i}=0 }}. \]
\DLE{$\sub$}
\item[$\sub$:] Let $(Y,a)\in \Se{M}$. Then
\EQN
a &\in& \S{N}(\ep[x\mto \downc{Y}]) \\
&=& \downc{\set{a_i\st \S{D_i}(\ep[x\mto \downc{Y}])=0}},
    \mbox{\ as $N\term Z$.}
\EEQN
Let $a\lep a_i$ and $\S{D_i}(\ep[x\mto\downc{Y}])=0$.
Then $\fl{\Sb{N_i}}\:(\downc{Y}) = 0$.
Hence $Y_i\sub\downc{Y}$, as $N_i\eq Y_i$.
So we get $(Y,a)\lep (Y_i,a_i)$.\\
Furthermore $\Se{C_i}=\S{C_i}(\ep[x\mto\downc{Y}])=0$.
\item[$\sup$:] Let $\Se{C_i}=0$.\\
We have $\fl{\Sb{N_i}}\:(\downc{Y_i})=0$, as $N_i\eq Y_i$.
Therefore $\S{D_i}(\ep[x\mto\downc{Y_i}])=0$.
As $N\term Z$, it is $a_i\in \S{N}(\ep[x\mto\downc{Y_i}])$.
Hence $(Y_i,a_i)\in \Se{M}$.
\EDLE

\medskip
\LP
{\bf 2)} Let $X\sub P_{n+1}(\s)$ be finite and consistent.
We construct $M\eq X$ by case analysis over $\s$.
\LP\BU $\s=\void$: $X=\emptyset$, $(\l x.\0)\eq X$.

\LP\BU $\s=\t+\r:$\\
If $X=\emptyset$, then $(\l x.\0)\eq X$.\\
Now let $a\in X$ for some $a\gep 0$.
Let $Y=\set{a\st (0,a)\in X} \sub P_n(\t)$.
By \ih there is some $N$ with $N\eq Y$.
Take $M=\l x.\case\:x\,N\,\1$.
It can be easily checked that $M\eq X$.\\
The case $a\in X$ for some $a\gep 1$ is similar.

\medskip
\LP\BU $\s=\t\p\r$:\\
Let $X_0=\set{a\st(0,a)\in X}\sub P_n(\t)$
and $X_1=\set{a\st(1,a)\in X}\sub P_n(\r)$.\\
There are $N_0\eq X_0$ and $N_1\eq X_1$ by induction hypothesis.\\
Let $M=\l x.\andf\:(N_0(\fst\:x))\:(N_1(\snd\:x))$.
We check easily that $M\eq X$.

\medskip
\LP\BU $\s=\t\f\r:$\\
If $X=\emptyset$, then $(\l x.\0)\eq X$.\\
Otherwise, let $X=\set{(Y_i,a_i)\st 1\leq i\leq k}$
be an enumeration of $X$.\\
Let $Y_i' = \set{\0\co b\st b\in Y_i} \sub \Cond(P_n(\t))$
for all $i$, it is finite and consistent.
By \ih there is $N_i\term Y_i'$ for all $i$.
Furthermore, by \ih there is $Q_i \eq {a_i}$ for all $i$.
We define
\[ M=\l x.\andf\:(Q_1(x N_1))(\andf\:(Q_2(x N_2))\ldots(Q_k(x N_k))) .\]
We check that $M\eq X$: Let $d\in D_\s$.\\
If $X\sub d$, then for all $i$:\\
$\fl{\Sb{Q_i}}\:(\fl{d}\:(\Sb{N_i})) =
  \fl{\Sb{Q_i}}\:(\fl{d}\:(\downc{Y_i})) = 0$,
as $a_i\in \fl{d}\:(\downc{Y_i})$. \\
Therefore $\fl{\Sb{M}}\:d = 0$.\\
If $d\ncon X$, then there is some $j$ with $d\ncon (Y_j,a_j)$,
\ie $\fl{d}\:(\downc{Y_j}) \ncon \set{a_j}$.
Therefore $\fl{\Sb{Q_j}}\:(\fl{d}\:(\Sb{N_j}))=1$,
and $\fl{\Sb{M}}\:d =1$.\\
Otherwise, $d\con(Y_i,a_i)$ for all $i$ and $(Y_j,a_j)\not\in d$
for some $j$.
Then $\fl{\Sb{M}}\:d=\bot$.

\medskip
We have now proved statements 1) and 2) for all $n$ and $\s$.
The lemma follows easily from 1):
If $d\in D_\s$ is finite, it has the form $d=\downc{X}$ with
$X\sub P_n(\s)$ for some $n$, $X$ finite and consistent.
There is a term $M$ with $M\term \set{\0\co a\st a\in X}$,
\ie $\Sb{M}=\downc{X}$.
\eproof

\medskip
\LP
{\bf Proof of the Full Abstraction Theorem:}\\
It remains to show for all $M,N\in \T_\s$:
If $M\suq N$, then $\Se{M}\sub \Se{N}$ for all $\ep$.
\LP
First suppose that $M$ and $N$ are closed terms.\\
Let $a\in \Se{M}$.
Define $f= \downc{(\set{a},0)} \in D_{\scriptsize \s\f\bool}$.
By the Definability Lemma, there is $P\in \Tc{\s\f\bool}$
with $\Sb{P}=f$.
$P[\:]$ serves as a context such that $PM$ and $PN$ are programs.\\
$0=\Sb{PM}= \Op{PM}\sub\Op{PN}=\Sb{PN}$,
therefore $a\in \Se{N}$.

\LP
Now let $M$ and $N$ be terms with their free variables in
$\set{x_1,\ldots,x_n}$.
We get $\l x_1\ldots x_n.M \suq \l x_1\ldots x_n.N$:
For all contexts $C[\:]$ apply the context
$C[\l x_1\ldots x_n.[\:]]$ to $M$ and $N$.
For the closed terms follows:
$\Se{\l x_1\ldots x_n.M} \sub \Se{\l x_1\ldots x_n.N}$ for all $\ep$.
Hence $\Se{M}\sub \Se{N}$ for all $\ep$.
\eproof

\section{Interdefinability of constants}

Our first observation is that \case\ can be defined from \pcase\
and \outz, \outi (see page \pageref{out} for the def.\ of \outz, \outi). We have
\[\S{\case} = \S{\l xyz.\pcase\:x\,(\pcase\:x\,(y\,(\outz\:x))\,\O)
                 (\pcase\:x\,\O\,(z\,(\outi\:x)))}.\]

In the preceding chapter we used the function $\andf: \bool\f\bool\f\bool$,
defined as $\andf=\l xy.\pcase\:x\,y\,\1$, to build defining terms for
all finite elements of the semantic model.
Now we will show that also \pcase\ (whose semantics is not finite) is
definable from \andf\ and the sequential constants.
Compare the definition of PCF's parallel conditional in terms of the
parallel or in \cite{Stoughton}.
We assume a constant $\andf: \bool\f\bool\f\bool$ with the semantics:
\[ (\Sb{\andf})00=0,\; (\Sb{\andf})1\bot=1,\; (\Sb{\andf})\bot 1=1.\]
Without loss of generality, we will define only
$\pcase_{\scriptsize \void,\void,\s}:\bool\f\s\f\s\f\s$ for all types $\s$,
and write simply $\pcase_\s$.
The general \pcase\ can be easily defined from this.

In order to cope with recursive types, we have to extend the inductive
definition of $\pcase_\s$ to general type expressions $\s$ (with free
type variables). Then we have to associate with each type variable $t$
of $\s$ some type $\t$ and a term variable $p:\bool\f\t\f\t\f\t$,
that stands for the $\pcase_\t$-function in its recursive definition.

So we are lead to define an operation $\Pcase(\te,\s)$ that produces terms
for \pcase-functions.
Its second argument is a type expression $\s\in\Tm$.
The first argument is a partial map $\te:V_T\f V$ from type variables
to term variables, with $\te(t)\neq \te(s)$ for $t\neq s$.
$\te$ is defined on a finite set of type variables that contains all
free variables of $\s$.
$\te(t)$ must be of the type $\bool\f\t\f\t\f\t$ for some type $\t$.
We associate with $\te$ the partial map $\ovte:V_T\f\Tmc$ defined
by $\ovte(t)=\t$ for $\te(t):\bool\f\t\f\t\f\t$.
$\Pcase(\te,\s)$ will be a term of type $\bool\f\ovte(\s)\f\ovte(\s)
\f\ovte(\s)$, where $\ovte$ is naturally extended to the substitution
of free type variables of type expressions.
$[\:]$ is the totally undefined map. The notation $\te[t\mto p]$ will
be used as for environments.

\LP
In the definition of \Pcase\ we use abbreviations for the following
function terms:
\vspace{-7mm}
\begin{tabbing}
\notf\  \= \kill\\
\ifz \> $: \bool\f\s\f\s\f\s$\\[\sma]
\ifz \> $= \l xyz.\case\:x\,(\l w.y)\,(\l w.z)$\\[\sma]
\notf \>$:\bool\f\bool$\\[\sma]
\notf \>$=\l x.\ifz\:x\,\1\,\0$ \\[\sma]
\orf \>$: \bool\f\bool\f\bool$\\[\sma]
\orf \>$=\l xy.\notf\:(\andf\:(\notf\:x)\,(\notf\:y))$\\[\sma]
\pcf \>$: \bool\f\bool\f\bool\f\bool$\\[\sma]
\pcf \>$=\l xyz. \orf\:(\orf\:(\andf\:x\,y)\,(\andf\:(\notf\:x)\,z))
       \:(\andf\:y\,z)$\\[\sma]
\> It is $\Sb{\pcf}=\Sb{\pcase_{\scriptsize \bool}}$.\\[\sma]
\sbf \>$: \t+\r\f\bool$\\[\sma]
\sbf \>$= \l x.\case\:x\,(\l y.\0)\,(\l y.\1)$
\end{tabbing}

\LP
$\Pcase(\te,\s)$ is defined by structural induction on the
type expression $\s$:
\vspace{-5mm}
\begin{tabbing}
$\Pcase(\te,\t\f\r)$ \= \kill\\
$\Pcase(\te,t)$ \> $= \te(t)$\\[\med]
$\Pcase(\te,\t+\r)$ \> $= \l x^{\sbool}y^{\ovte(\t+\r)}z^{\ovte(\t+\r)}.
 \ifz$ \= $(\pcf\:x\,(\sbf\:y)\,(\sbf\:z))$\\[\sma]
\> \> $(\0\:(\Pcase(\te,\t)\:x\,(\outz\:y)\,(\outz\:z)))$\\[\sma]
\> \> $(\1\:(\Pcase(\te,\r)\:x\,(\outi\:y)\,(\outi\:z)))$\\[\med]
$\Pcase(\te,\t\p\r)$ \> $= \l x^{\sbool}y^{\ovte(\t\p\r)}z^{\ovte(\t\p\r)}.
 ($ \=
 $\Pcase(\te,\t)\:x\,(\fst\:y)\,(\fst\:z),$\\[\sma]
 \> \> $\Pcase(\te,\r)\:x\,(\snd\:y)\,(\snd\:z))$\\[\med]
$\Pcase(\te,\t\f\r)$ \> $= \l x^{\sbool}y^{\ovte(\t\f\r)}z^{\ovte(\t\f\r)}
  w^{\ovte(\t)}. \Pcase(\te,\r)\:x\,(y\,w)(z\,w)$\\[\med]
$\Pcase(\te,\m t.\t)$ \> $=Y_\pi(\l p^\pi.\Pcase(\te[t\mto p^\pi],\t)),$\\[\sma]
\> where $\pi=\bool\f\ovte(\m t.\t)\f\ovte(\m t.\t)\f\ovte(\m t.\t)$,\\[\sma]
\> and $p^\pi$ denotes the first variable in $V^\pi$ that is not in the
   image of $\te$ \\[\med]
$\Pcase(\te,\void)$ \> $= \O$
\end{tabbing}

\LP
It is easy to show by induction that 
$\Pcase(\te,\s): \bool\f\ovte(\s)\f\ovte(\s)\f\ovte(\s)$.\\
In the case of the recursive type expression we have
\EQN
\Pcase(\te[t\mto p^\pi],\t) &:& \bool\f\r\f\r\f\r \\
\mbox{with\ } \r &=& \ov{\te[t\mto p^\pi]}(\t)\\
&=& (\ovte[t\mto \ovte(\m t.\t)])(\t)\\
&=& \ovte([\:][t\mto \m t.\t](\t))\\
&\approx& \ovte(\m t.\t),
\EEQN
so $\Pcase(\te[t\mto p^\pi],\t): \pi$,
therefore $\Pcase(\te,\m t.\t): \pi$.
\LP
$\Pcase(\te,\s)$ has the free variables $\te(t)$ for all $t$ free in $\s$.

\DEFI
Let $f\in D_{\sbool\f\t\f\t\f\t}$ for some type $\t$.\\
We say that $f$ {\de approximates} the function \spcase\ {\de to level}
$n$, $\app_n(f)$, iff $f\,c\,a\,b \sup(\spcase\:c\,a\,b)|_n$
for all $c\in D_{\sbool}$ and $a,b\in D_\t$.
\EDEFI
\LEMMA
Let $\te,\s$ be admissible arguments in $\Pcase(\te,\s)$, as described
above.
Let $n\geq 0$ and $\ep$ be an environment with
$\app_n(\ep(\te(t)))$ for all $t$ free in $\s$.\\
Then for $f=\Se{\Pcase(\te,\s)}$ we have $\app_n(f)$.\\
If $\s$ is not of the form $\m t_1\ldots \m t_m.t$,
with $m\geq 0$,
$t$ a type variable and $t\neq t_i$ for all $i$, then $\app_{n+1}(f)$.
\ELEMMA
\proof by structural induction on $\s$.
\LP
\BU $\s=t$: $f=\ep(\te(t))$, hence $\app_n(f)$.
\LP
\BU $\s=\t+\r$:\\
We show $\app_{n+1}(f)$, \ie
$f\,c\,a\,b \sup (\spcase\:c\,a\,b)|_{n+1}$ for all $c\in D_{\sbool}$,
$a,b\in D_{\ovte(\s)}$.
\DLE{3)}
\item[1)] $c=\bot:$\\
The case $a\cap b=\bot$ is clear.\\
Now let $a=0a'$, $b=0b'$.
\EQN
f\,\bot\,(0a')\,(0b') &=& 0\,((\Se{\pcase(\te,\t)})\,\bot\,a'\,b')\\
&\sup& 0\,((a'\cap b')|_n), \mbox{\ by \ih}\\
&=& (a\cap b)|_{n+1}\\
&=& (\spcase\:c\,a\,b)|_{n+1}
\EEQN
The case $a=1a'$, $b=1b'$ is analogous.
\item[2)] $c=0$:\\
The case $a=\bot$ is clear.\\
Now let $a=0a'$.
\EQN
f\,0\,(0a')\,b &=& 0\,((\Se{\Pcase(\te,\t)})\,0\,a'\,((\Sb{\outz})\,b))\\
&\sup& 0\,(a'|_n), \mbox{\ by \ih}\\
&=& a|_{n+1}\\
&=& (\spcase\:c\,a\,b)|_{n+1}
\EEQN
The case $a=1a'$ is analogous.
\item[3)] $c=1$ is analogous to $c=0$.
\EDLE

\medskip
\LP
\BU $\s=\t\p\r:$\\
We show $\app_{n+1}(f)$.
For all $c\in D_{\sbool}$, $a_1,b_1\in D_{\ovte(\t)}$
and $a_2,b_2\in D_{\ovte(\r)}$ we have:
\EQN
f\,c\,(\spair\:a_1 a_2)\,(\spair\:b_1 b_2) &=&
  \spair\:((\Se{\Pcase(\te,\t)})\,c\,a_1\,b_1)
  ((\Se{\Pcase(\te,\r)})\,c\,a_2\,b_2)\\
&\sup& \spair\:(\spcase\:c\,a_1\,b_1)|_n\,(\spcase\:c\,a_2\,b_2)|_n,
   \mbox{\ by induction hyp.}\\
&=& (\spair\:(\spcase\:c\,a_1\,b_1)\,(\spcase\:c\,a_2\,b_2))|_{n+1}\\
&=& (\spcase\:c\,(\spair\:a_1\,a_2)\,(\spair\:b_1\,b_2))|_{n+1}
\EEQN

\LP \BU $\s=\t\f\r:$\\
We prove $\app_{n+1}(f)$.
Let $c\in D_{\sbool}$ and $a,b\in D_{\ovte(\s)}$.\\
$ f\,c\,a\,b = \Pr(d\in D_{\ovte(\t)} \mto
  (\Se{\Pcase(\te,\r)})\,c\,(a\,d)\,(b\,d))$.\\
Let $(X,r)\in (\spcase\:c\,a\,b)|_{n+1}$. Then
\EQN
r &\in& ((\spcase\:c\,a\,b)\,(\downc{X}))|_n \\
&=& (\spcase\:c\,(a\,(\downc{X}))\,(b\,(\downc{X})))|_n \\
&\sub& (\Se{\Pcase(\te,\r)})\,c\,(a\,(\downc{X}))\,(b\,(\downc{X})),
  \mbox{\ by \ih}
\EEQN
Hence $(X,r)\in f\,c\,a\,b$.

\medskip
\LP\BU $\s=\m t.\t:$
\DLE{1.2)}
\item[1)]
We assume that $\t$ is {\em not} of the form 
$\m t_1\ldots \m t_m.s$ with $m\geq 0$, $s$ a type variable, $s\neq t$,
and $s\neq t_i$ for all $i$.
We have to show $\app_{n+1}(f)$.
\item[1.1)]
We assume $\t=\m t_1\ldots \m t_m.t$.\\
Then $\ovte(\s)=\s \approx \void$,
hence $f\in D_{\scriptsize \bool\f\void\f\void\f\void}$ 
and $\app_{n+1}(f)$.
\item[1.2)]
Otherwise, $\t$ is not of the form
$\m t_1\ldots \m t_m.s$ with $m\geq 0$,
$s$ a type variable and $s\neq t_i$ for all $i$.\\
We have $f= \bigcup_{i\geq 0} g^i\bot$
with $g= \fl{\Se{\l p.\Pcase(\te[t\mto p],\t)}}$.\\
We show by induction on $i$ that $\app_i(g^i\bot)$
for $0\leq i\leq n+1$:\\
$\app_0(g^0\bot)$ is trivial.\\
Induction step: We assume $\app_i(g^i\bot)$ for some $i\leq n$.\\
$g^{i+1}\bot = g(g^i\bot) =
 \S{\Pcase(\te[t\mto p],\t)}(\ep[p\mto g^i\bot])$.\\
By the general \ih (for the type expression $\t$) we get
$\app_{i+1}(g^{i+1}\bot)$.
Especially we have $\app_{n+1}(g^{n+1}\bot)$, hence $\app_{n+1}(f)$.
\item[2)]
We assume $\t=\m t_1\ldots \m t_m.s$ with $m\geq 0$,
$s$ a type variable, $s\neq t$,
and $s\neq t_i$ for all $i$.\\
Then $f=\Se{\Pcase(\te,s)} = \ep(\te(s))$, so $\app_n(f)$.
\EDLE

\LP\BU $\s=\void:$ Trivial.
\eproof

\THEOREM
Let $\te, \s$ be admissible arguments in $\Pcase(\te,\s)$, as described 
above.
Let $\ep$ be an environment with 
$\ep(\te(t))=\Sb{\pcase_{\ovte(t)}}$ for all $t$ free in $\s$.
Then $\Se{\Pcase(\te,\s)}=\Sb{\pcase_{\ovte(\s)}}$.
Especially for all types $\s$ we have:
$\Sb{\Pcase([\:],\s)}=\Sb{\pcase_\s}$.
\ETHEOREM
\proof
$\Se{\Pcase(\te,\s)}\sup \Sb{\pcase_{\ovte(\s)}}$ follows from the
preceding lemma.\\
Now let $f=\Se{\Pcase(\te,\s)}$.
We show $f\sub \Sb{\pcase_{\ovte(\s)}}$ by structural induction on $\s$:

\LP\BU $\s=t:$ $f=\ep(\te(t))= \Sb{\pcase_{\ovte(t)}}$.

\medskip
\LP\BU $\s=\t+\r:$\\
We show $f\,c\,a\,b \sub \spcase\:c\,a\,b$ for all $c\in D_{\sbool}$
and $a,b\in D_{\ovte(\s)}$.
\DLE{3)}
\item[1)] $c=\bot:$\\
For $a\cap b=\bot$ it is $f\,\bot\,a\,b = \bot$.\\
Now let $a=0a'$, $b=0b'$.
\EQN
f\,\bot\,(0a')\,(0b') &=& 0\,((\Se{\Pcase(\te,\t)})\,\bot\,a'\, b') \\
&\sub& 0\,(\spcase\,\bot\,a'\,b'),\mbox{\ by \ih}\\
&=& \spcase\,\bot\,a\,b
\EEQN
The case $a=1a'$, $b=1b'$ is analogous.
\item[2)] $c=0:$\\
For $a=\bot$ it is $f\,0\,\bot\,b =\bot$.\\
Now let $a=0a'$.
\EQN
f\,0\,(0a')\,b &=& 0\,((\Se{\Pcase(\te,\t)})\,0\,a'\,((\Sb{\outz})b)) \\
&\sub& 0\,(\spcase\:0\,a'\,((\Sb{\outz})b)),\mbox{\ by \ih}\\
&=& \spcase\:0\,a\,b
\EEQN
The case $a=1a'$ is analogous.
\item[3)] $c=1$ is analogous to $c=0$.
\EDLE

\medskip
\LP\BU $\s=\t\p\r:$\\
For all $c\in D_{\sbool}$, $a_1,b_1\in D_{\ovte(\t)}$
and $a_2,b_2\in D_{\ovte(\r)}$:
\EQN
f\,c\,(\spair\:a_1 a_2)\,(\spair\:b_1 b_2) &=&
  \spair\:((\Se{\Pcase(\te,\t)})\,c\,a_1\,b_1)
  ((\Se{\Pcase(\te,\r)})\,c\,a_2\,b_2)\\
&\sub& \spair\:(\spcase\:c\,a_1\,b_1)\,(\spcase\:c\,a_2\,b_2),
   \mbox{\ by induction hyp.}\\
&=& \spcase\:c\,(\spair\:a_1\,a_2)\,(\spair\:b_1\,b_2)
\EEQN

\LP\BU $\s=\t\f\r:$\\
For all $c\in D_{\sbool}$, $a,b\in D_{\ovte(\s)}$ and $d\in D_{\ovte(\t)}$:
\EQN
f\,c\,a\,b\,d &=& (\Se{\Pcase(\te,\r)})\,c\,(a\,d)\,(b\,d) \\
&\sub& \spcase\:c\,(a\,d)\,(b\,d),\mbox{\ by \ih}\\
&=& \spcase\:c\,a\,b\,d
\EEQN

\LP\BU $\s=\m t.\t:$\\
$f$ is the least fixed point of $g=\fl{\Se{\l p.\Pcase(\te[t\mto p],\t)}}$.\\
Let $d=\Sb{\pcase_{\ovte(\s)}}$. Then
\EQN
g\,d &=& \S{\Pcase(\te[t\mto p],\t)}(\ep[p\mto d])\\
&\sub& \Sb{\pcase_\r}\mbox{\ with\ }
   \r=\overline{\te[t\mto p]}(\t) \approx \ovte(\s),\mbox{\ by \ih}\\
&=& d.
\EEQN
Therefore $f\sub d$.

\LP\BU $\s=\void:$ Trivial.
\eproof

\section{Conclusion}

We have given the syntax and reduction relation of a recursively typed
$\l$-calculus with a parallel conditional \pcase\ on all types.
The calculus was proved to be confluent, with the aid of a general result
on the confluence of the $\l$-calculus with algebraic term rewriting rules.
Our reduction relation simply defines the reduction of a redex in any
context. It remains to define a reduction strategy that effectively
finds the normal form approximations of a term.
Such a strategy cannot prescribe deterministically which redex to reduce,
as we have the parallel \pcase.
Instead, it should give for every term a set of its outermost redexes
to be reduced in the next reduction steps.
Such a strategy could be given for general algebraic term rewriting rules.

We unfolded the recursive types to (possibly infinite) type trees
and interpreted these type trees as prime systems.
With this interpretation of types, we gave a denotational semantics of terms.
The Approximation Theorem was the key result on the strength of reduction
with respect to the denotational semantics: 
The semantics of a term equals the limit of the semantics of its
normal form approximations.
From this followed the adequacy of the semantics with respect to the
observation of Boolean values: 
If the semantics of a program is $0$ or $1$, then the program reduces
to this value.
Furthermore, we showed full abstraction of the semantics.
To achieve this, the syntax must contain a parallel function like
\pcase\ or \andf. 
These functions are definable from each other, so a calculus with the same
expressive power could be given with reduction rules for \andf\ instead
of \pcase.
The same expressive power means that the same elements of the semantic
model are definable in both calculi.
The semantic model corresponds to the observation of Boolean values,
as we have seen. There are other operational, intensional properties of the
original \pcase\ that are not valid for the \pcase-function defined from \andf,
\eg the reduction $\pcase\:\0\,M N\rast M$.
The proofs of confluence and of the Approximation Theorem would
be (slightly) easier for a calculus with \andf.
Nevertheless, we preferred to make these investigations with a
\pcase-calculus.

\bigskip
\LP
{\bf Acknowledgements:} I thank Reinhold Heckmann for carefully reading
a draft of this paper.

\bibliographystyle{alpha}

\end{document}